\documentclass[aps,prl,reprint,preprintnumbers,groupedaddress,nofootinbib,floatfix,
  superscriptaddress,nolongbibliography]{revtex4-2}
\usepackage{amsmath,amssymb,graphicx,xcolor}
\usepackage[colorlinks=true,citecolor=red,linkcolor=black,urlcolor=black]{hyperref}
\usepackage{xspace,orcidlink}
\newcommand{\neff}{\ensuremath{N_{\rm eff}}\xspace}
\newcommand{\yp}{\ensuremath{Y_{\rm P}}\xspace}
\providecommand{\dh}{}
\renewcommand{\dh}{\ensuremath{{\rm D/H}}\xspace}
\newcommand{\Dfive}{\ensuremath{10^5{\rm D/H}}\xspace}
\newcommand{\Trh}{\ensuremath{T_{\rm RH}}\xspace}
\newcommand{\Trhlow}{\ensuremath{T_{\rm RH,low}}\xspace}
\newcommand{\mX}{\ensuremath{m_X}\xspace}
\newcommand{\tX}{\ensuremath{\tau_X}\xspace}
\newcommand{\MeV}{\ensuremath{{\rm MeV}}\xspace}
\newcommand{\orcid}[1]{\,\orcidlink{#1}}
\newcommand{\nudsmc}{\ensuremath{\nu}\textsc{Dsmc}\xspace}
\newif\ifshowauthorcomments
\showauthorcommentsfalse

\begin{document}
\preprint{CERN-TH-2026-271}
\title{Late reheating in the era of precision cosmology}
\author{Kensuke~Akita\orcid{0000-0002-5919-9347}}
\email{kensuke.akita.e2@tohoku.ac.jp}
\affiliation{Department of Physics, Tohoku University, Sendai, Miyagi 980-8578, Japan}
\affiliation{Department of Physics, The University of Tokyo, Bunkyo-ku, Tokyo 113-0033, Japan}
\author{Oleksii~Ihnatenko\orcid{0009-0004-6596-6137}}
\email{oleksii.ihnatenko@kit.edu}
\affiliation{Taras Shevchenko National University of Kyiv, 64 Volodymyrs'ka str., Kyiv 01601, Ukraine}
\affiliation{Institute for Astroparticle Physics, Karlsruhe Institute of Technology, Hermann-von-Helmholtz-Platz 1, 76344 Eggenstein-Leopoldshafen, Germany}
\author{Maksym~Ovchynnikov\orcid{0000-0001-7002-5201}}
\email{maksym.ovchynnikov@cern.ch}
\affiliation{Theoretical Physics Department, CERN, 1 Esplanade des Particules, CH-1211 Geneva 23, Switzerland}
\hypersetup{pdftitle={Late reheating in the era of precision cosmology},
 pdfauthor={Kensuke Akita, Oleksii Ihnatenko, Maksym Ovchynnikov}}

\begin{abstract}
We study the minimum cosmic reheating temperature allowed by recent measurements of primordial nuclear abundances and of the cosmic microwave background. We find that reheating by decays of a light particle into electrons and neutrinos is compatible with both at temperatures around $0.4\,\MeV$, an order of magnitude below the bound for purely electromagnetic channels. This scenario has implications for cosmology and laboratory searches. For example, nearly standard neutrino radiation can coexist with a smaller active neutrino mass density at unchanged physical masses, by up to a factor of five for normal ordering. The weaker suppression of clustering changes matter power at the percent level, which forthcoming galaxy surveys and lensing measurements may probe. Late reheating also suppresses sterile neutrino production at mixings within the projected reach of laboratory experiments.
\end{abstract}
\maketitle

\begin{figure*}[t!]
 \centering
 \includegraphics[width=0.485\textwidth]{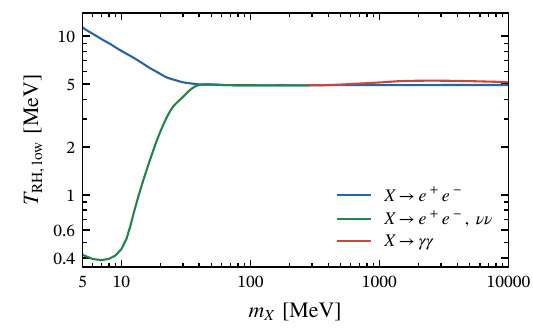}\hfill
 \includegraphics[width=0.485\textwidth]{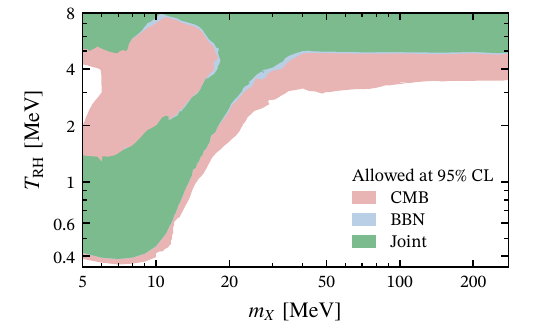}
 \caption{Cosmological constraints on scalar reheating. Left: the bound on the lowest possible reheating temperature $\Trhlow$, Eq.~\eqref{eq:minimum}, versus the mass $m_X$ of the scalar reheater. Blue (red): dominant decays into $e^+e^-$ (photons). Green: mixed decays into $e^+e^-$ and neutrinos of different flavors. The separation between the blue and red curves at $m_X\gtrsim2m_{\pi^\pm}$ is caused by subdominant decays producing charged pions in the $\gamma\gamma$ case, evaluated at fixed electromagnetic plasma evolution. Right: for mixed decays, regions compatible with CMB-SPA~\cite{SPT-3G:2025bzu}, primordial abundances (BBN)~\cite{Yeh:2026lbt,Barbieri:2025moq}, and their joint likelihood, using Eq.~\eqref{eq:joint-region} for each likelihood separately. Baryon density and decay fractions are profiled within the flavor families specified in SuM. All calculations assume the initial reheater abundance given in the text.}
 \label{fig:minima}
\end{figure*}

Reheating creates the thermal plasma of the hot Big Bang. This happens at the end of inflation or after a later era in which a long-lived massive particle dominates the energy density until it decays~\cite{Coughlan:1983ci,Moroi:1999zb}. The initial temperature of this plasma determines which relics, such as dark matter and the baryon asymmetry, can be produced thermally~\cite{Giudice:2000ex}.

Late reheating has been considered for its implications for sterile neutrinos and axions, among others~\cite{Gelmini:2004ah,Visinelli:2009kt,Hasegawa:2020ctq}. They depend strongly on how long this particle can dominate. The radiation era may then begin only shortly before Big Bang nucleosynthesis (BBN). Near the MeV scale, however, weak interactions may no longer establish thermal equilibrium for neutrinos, changing both light element production and the cosmic neutrino background~\cite{Kawasaki:1999na,Kawasaki:2000en}. This duration therefore depends on whether its decays produce neutrinos directly.

We define the reheating temperature $\Trh$ through the vacuum width $\Gamma_X$ and lifetime $\tX$ of the long-lived particle:
\begin{equation}
 \begin{aligned}
 \Gamma_X&\equiv\tX^{-1}=3H_{\rm rad}(\Trh),\\
 \tX&\simeq0.49\,{\rm s}\left(\frac{\MeV}{\Trh}\right)^2,
 \end{aligned}
 \label{eq:Trh}
\end{equation}
where $H_{\rm rad}$ is the Hubble rate for equilibrium radiation with $g_*=10.75$ effective relativistic degrees of freedom. BBN and the cosmic microwave background (CMB) constrain how low $\Trh$ can be~\cite{Pitrou:2018cgg,Planck:2018nkj}.

Most previous studies considered reheating by decays into electromagnetic or hadronic species~\cite{Kawasaki:1999na,Kawasaki:2000en,Ichikawa:2005vw,deSalas:2015glj,Hasegawa:2019jsa,Barbieri:2025moq}. Electromagnetic decays rely on weak reactions in the plasma to populate neutrinos, while hadronic channels can strengthen BBN constraints by converting protons into neutrons~\cite{Kawasaki:2000en,Hasegawa:2019jsa,Cline:2026lba}. Depending on the final states and how conservatively the observations are interpreted, the lower limits reach approximately $5$ to $6\,\MeV$~\cite{Hasegawa:2019jsa,Barbieri:2025moq}.

Ref.~\cite{Hannestad:2004px} studied decays into both electromagnetic particles and neutrinos, which we call mixed decays, and found allowed regions near $\Trh=1\,\MeV$ for equal branching fractions into the three neutrino flavors. However, the analysis omitted neutrino oscillations, neutrino-neutrino collisions, inverse reactions, and Pauli blocking of the daughter states. These approximations can change whether the reported regions are viable. The analysis also used the abundance, CMB, and galaxy survey data of 2004, far less precise than present measurements~\cite{Yeh:2026lbt,SPT-3G:2025bzu,Barbieri:2025moq}.

\begingroup\clubpenalty=10000
In this work, we show that mixed decays with unequal branching fractions into the neutrino flavors satisfy the joint CMB and BBN constraints for reheating temperatures down to $\Trh\simeq0.4\,\MeV$ (Fig.~\ref{fig:minima}). Close to this bound, $\neff$ and the yields of light elements remain near their standard values, while fewer relic neutrinos carry higher mean momenta.

The Supplemental Material (SuM) gives the necessary technical details.
\par\endgroup
\paragraph{Reheating scenarios.}
We consider a scalar $X$ of mass $\mX$, called the reheater, that is initially at rest and dominates the total energy density.\footnote{Pseudoscalar and vector reheaters are also possible, with different quantitative bounds. In anomaly-free gauge extensions of the Standard Model, however, charge assignments often restrict the relative electron and neutrino widths of a vector mediator~\cite{Ilten:2018crw,Kamada:2015era}.} At a temperature of $20\,\MeV$, we take $\rho_X/\rho_{\rm rad}=7\times10^5\,m_X/\MeV$, which corresponds to a fixed number of reheaters per unit entropy. We have checked that Fig.~\ref{fig:minima} does not visibly change for smaller values of this ratio. The actual Hubble rate is $H=(8\pi G\rho_{\rm tot}/3)^{1/2}$, with Newton's constant $G$ and total energy density $\rho_{\rm tot}$ including $X$. Initially, $\rho_X$ makes $\Gamma_X/H\ll1$. While $X$ dominates, $H$ exceeds $H_{\rm rad}(T)$, and its decays add entropy.

Purely electromagnetic (EM) late reheating heats the photon-electron plasma while weak reactions fail to fully populate neutrinos. We denote the EM plasma temperature by $T$ throughout. The neutrino deficit lowers the effective number of neutrinos, defined after reheating and $e^+e^-$ annihilation by
\begin{equation}
 \neff=\frac{8}{7}\left(\frac{11}{4}\right)^{4/3}
 \frac{\rho_\nu+\rho_{\rm DR}}{\rho_\gamma},
 \label{eq:Neff}
\end{equation}
where $\rho_\nu$, $\rho_\gamma$, and $\rho_{\rm DR}$ are the energy densities of active neutrinos, photons, and additional dark radiation. Depletion of electron neutrinos weakens neutron-proton conversion, while reduced $\neff$ slows expansion. At sufficiently low reheating temperatures, the neutron-to-proton ratio and hence the helium abundance increase, because the missing $\nu_e$ and $\bar\nu_e$ outweigh the smaller Hubble rate~\cite{Kawasaki:2000en}. During nuclear burning, the extra time also lets more deuterium be consumed, which decreases the deuterium-to-hydrogen number ratio $\dh$. Evading the EM reheating bound therefore requires recovering both the radiation density and the yields of light elements.

Adding dark radiation to electromagnetic reheating can restore the standard $\neff$. However, it speeds up the expansion, while the active neutrinos responsible for neutron conversion remain depleted, and both effects raise the helium abundance. Consequently, this addition does not substantially lower the reheating bound.

Enhanced neutrino self-interactions can establish kinetic equilibrium among neutrinos, but they conserve the total neutrino number and energy. Neutrinos thus still gain both only through weak reactions with the EM plasma, which remain too slow during late reheating, so the depletion persists.

We therefore turn to mixed decays, which produce neutrinos directly. For reheater masses $m_X\gtrsim40\,\MeV$, however, the injected energetic neutrinos raise the neutron-to-proton ratio and strengthen the lower bound on $\Trh$. Self-interactions cannot remove this effect, because they conserve neutrino number and energy and thus leave unchanged the mean neutrino momentum increased by these decays. Lower reheating temperatures therefore require lighter reheaters, whose neutrinos are softer.

\paragraph{Neutrino spectra and light elements.}
We study decays into electrons, photons, and neutrinos. Additional hadronic decays would enhance neutron production and generally strengthen the lower bound on $\Trh$~\cite{Boyarsky:2020dzc,Hasegawa:2019jsa}.

The decay pattern is specified by the vacuum branching fractions into electron pairs, $B_{ee}$, and into two active neutrinos of flavor $\alpha=e,\mu,\tau$, $B_{\nu_\alpha}$, which obey\footnote{Decays into two photons have distinct inverse reactions and Bose factors; the bounds for light reheaters therefore specifically concern the channels in Eq.~\eqref{eq:branching}.}
\begin{equation}
 B_{ee}+B_{\nu_e}+B_{\nu_\mu}+B_{\nu_\tau}=1,
 \qquad \Gamma_c^{\rm vac}=B_c\Gamma_X,
 \label{eq:branching}
\end{equation}
where $c$ labels the four channels and $\Gamma_c^{\rm vac}$ is the corresponding vacuum partial width. The neutrino couplings are diagonal in the charged lepton flavor basis, and decays inject equal populations of active neutrinos and antineutrinos. Such direct injection can lower the bound for both Dirac and Majorana neutrinos; its value depends on the particle model.

Injected electron-positron pairs rapidly thermalize through electromagnetic interactions. In the CP-even scalar model used for our numerical bounds, we include the thermal electron mass in the kinematics of decays and inverse reactions.

We consider reheater masses up to $10\,$GeV; heavier ones should not lower the minimum reheating temperature. For electromagnetic decays, small probabilities for producing metastable mesons such as $\pi^\pm$ generally arise at $m_X\gtrsim2m_{\pi^\pm}$. Examples include $X\to\gamma\pi^+\pi^-$ and, above the threshold for an additional electron pair, $X\to e^+e^-\pi^+\pi^-$. Even tiny decay probabilities into charged pions can raise the neutron-to-proton ratio and strengthen the late reheating bounds~\cite{EscuderoAbenza:2025tsi}. Fig.~\ref{fig:minima} includes this contribution, which weakens above a few GeV as their yield per unit of released energy falls (Sec.~\ref{sec:sm-pions}).

Electron neutrino and antineutrino spectra determine the neutron-to-proton ratio through processes such as $\nu_e+n\leftrightarrow p+e^-$ and $\bar\nu_e+p\leftrightarrow n+e^+$. This ratio sets the primordial helium abundance $\yp=4n_{{}^4\mathrm{He}}/n_b$, where $n_b$ is the baryon number density.

Decays inject neutrinos with $p\simeq\mX/2$. Depending on the plasma temperature during injection, this value can lie below, near, or above the mean for thermal neutrinos. Predicting $\neff$ and nuclear abundances therefore requires resolving the momentum distributions. The flavors cannot be treated as a single fluid either: fast oscillations do not generically equalize their populations, especially when $B_{\nu_e}$, $B_{\nu_\mu}$, and $B_{\nu_\tau}$ differ.

We calculate this coupled evolution with the Neutrino Direct Simulation Monte Carlo method (\nudsmc)~\cite{Ovchynnikov:2024rfu,Ovchynnikov:2024xyd,Ihnatenko:2025kew,Akita:2026xwu}, following the neutrino distributions, cold and moving $X$ populations, electromagnetic temperature, and scale factor. The reactions include scattering on electrons and neutrinos, pair annihilation, decays, and inverse decays, with quantum statistical factors and phase-averaged flavor oscillations. Our BBN framework uses \textsc{Primat} nuclear rates~\cite{Pitrou:2018cgg,Pitrou:2020etk} and finite nucleon mass corrections to weak neutron-proton conversion~\cite{Akita:2026xwu}.

Comparisons with quasi-classical Boltzmann and quantum kinetic calculations give close agreement for electromagnetic reheating and subpercent differences in $\neff$ for mixed decays, with and without inverse reactions~\cite{Akita:2026xwu}.

\paragraph{Joint cosmological constraints.}
We combine the CMB-SPA likelihood from Planck, ACT, and SPT
observations~\cite{SPT-3G:2025bzu} with measured primordial
abundances of light elements and the nuclear uncertainties of their
predictions. The CMB data constrain the physical baryon density
$\omega_b$, $\neff$, and the helium mass fraction
$Y_{\rm CMB}$ used in recombination, with correlated errors.
Reheating changes the standard BBN relation of $Y_{\rm CMB}$ to
$\omega_b$ and $\neff$, so we use this likelihood with free
$\neff$ and $Y_{\rm CMB}$ and obtain $Y_{\rm CMB}$ from the
calculated $\yp$.
CMB-SPA treats the relic neutrinos as thermal, with
$\sum_im_i=0.06\,$eV. Late reheating breaks this assumption: decays
predominantly into $\nu_e$ deplete the heaviest eigenstate, and those
into $\nu_\tau$ enrich it. These changes affect the CMB mainly through
the present neutrino mass density, which suppresses lensing. To test the
assumption, we refit the data with the calculated distributions for
histories near the boundary of the allowed region (see SuM). The CMB
$\chi^2$ shifts by $-0.3$ to $+0.5$ relative to thermal neutrinos.
Below the lower bound on $\Trh$, the joint $\chi^2$ rises steeply
because of helium, so this limit moves by less than $0.001\,\MeV$.
We therefore keep thermal neutrinos in the CMB likelihood.

We adopt the measurements of primordial helium and deuterium abundances from Refs.~\cite{Yeh:2026lbt} and~\cite{Barbieri:2025moq}, respectively:
\begin{equation}
 (\yp,\Dfive)_{\rm obs}
 =(0.2458\pm0.0013,\;2.547\pm0.025),
 \label{eq:bbn_data}
\end{equation}
with one-standard-deviation uncertainties. The predicted yields depend on the baryon-to-photon number ratio, which entropy release lowers during reheating; $\omega_b$ fixes its final value.

Unlike Ref.~\cite{Barbieri:2025moq}, we omit DESI baryon acoustic oscillation data and other galaxy surveys. DESI and CMB measurements mildly disagree on the expansion history within $\Lambda$CDM~\cite{DESI:2025DR2}; excluding these additional constraints gives conservative reheating bounds without assuming their consistency.

The function $\mathcal L_{\rm joint}$ multiplies the CMB and
abundance likelihoods and averages the product over the nuclear
uncertainties of the predicted $\yp$ and $\dh$.
At each reheater mass $m_X$ and temperature $\Trh$, we minimize
$\chi^2=-2\ln\mathcal L_{\rm joint}+\mathrm{constant}$ over
$\omega_b$ and the branching fractions $\boldsymbol B$ within
a decay class $\mathcal B$. Electron-only decays have $B_{ee}=1$;
mixed decays allow electron and neutrino channels. The lower
temperature limits in Fig.~\ref{fig:minima} (left) are defined
relative to the best fit at the same mass and within the same
decay class:
\begin{align}
 q_{m_X}(\Trh;\mathcal B)&=\min_{\boldsymbol B\in\mathcal B,\omega_b}\chi^2
 -\chi^2_{\min}(m_X;\mathcal B),\notag\\
 \Trhlow(m_X;\mathcal B)&=\inf\{\Trh:q_{m_X}(\Trh;\mathcal B)\leq4\}.
 \label{eq:minimum}
\end{align}
The reference $\chi^2_{\min}(m_X;\mathcal B)$ is also minimized
over $\Trh$. We use $q_{m_X}\leq4$ for the asymptotic
two-standard-deviation limit in one parameter.
For the 95\% regions in $(m_X,\Trh)$ shown in
Fig.~\ref{fig:minima} (right), we profile over $\omega_b$ and
the decay fractions and use the minimum over masses,
temperatures, and decay patterns as the reference:
\begin{equation}
 \Delta\chi^2\equiv\chi^2-\chi^2_{\rm global}\leq5.99.
 \label{eq:joint-region}
\end{equation}
The $m_X=6\,\MeV$, $\Trh=0.42\,\MeV$ example in Table~\ref{tab:examples} has
$\Delta\chi^2\simeq1.78$ and satisfies both criteria,
Eqs.~\eqref{eq:minimum} and~\eqref{eq:joint-region}.

With \textsc{Primat} rates, standard BBN underproduces deuterium by about $2.3\sigma$ at its joint best-fit baryon density, including observational and nuclear uncertainties. Late reheating can alter $\dh$: for nearly thermal neutrino spectra at fixed final $\omega_b$, $\Delta(\dh)/(\dh)_{\rm SBBN}\simeq0.11\,\Delta\neff$~\cite{Akita:2026xwu}. Mixed decays further affect $\dh$ through the expansion rate, the entropy release, and neutron conversion by non-thermal $\nu_e$ and $\bar\nu_e$. For a spatially homogeneous reheater, the histories compatible with the joint constraints give $\dh$ close to standard BBN and do not remove the deuterium deficit. Density perturbations of $X$, which grow while it dominates, make its decays heat the plasma unevenly, so the baryon-to-photon ratio varies in space. Because $\dh$ scales roughly as its $-1.6$ power, averaging over such regions raises $\dh$ and could resolve this tension~\cite{Barrow:2018yyg,Bagherian:2025puf} (see SuM).

\paragraph{Lowest possible reheating temperatures.}
Decays into electrons and neutrinos give the lowest reheating temperatures for masses of a few to ten MeV (Fig.~\ref{fig:minima}, left).

For heavy reheaters, energetic neutrinos efficiently transfer energy to the EM plasma through annihilation and scattering. The plasma rapidly redistributes this energy among many softer particles, which return less of it to neutrinos, so $\neff$ can decrease~\cite{Hannestad:2004px,Akita:2026xwu}. The energetic electron neutrinos and antineutrinos also have nearly equal capture cross sections on neutrons and protons. Since protons are more abundant, these reactions produce extra neutrons and hence helium. Together, these effects prevent heavy mixed decays from substantially weakening the EM reheating bound.

Lowering the reheater mass reduces the injected neutrino energy and the subsequent energy transfer to the EM plasma. Direct neutrino injection can then compensate the neutrino deficit of EM reheating even at smaller $\Trh$. A simple estimate of the required total neutrino branching fraction $B_\nu=\sum_\alpha B_{\nu_\alpha}$ follows from energy balance. For instantaneous decay into a relativistic electron plasma, neglecting medium corrections and subsequent energy exchange, entropy conservation gives
\begin{equation}
 \frac{B_\nu}{B_{ee}}\simeq\frac{7}{22}\neff.
 \label{eq:branch-balance}
\end{equation}
Near-standard $\neff$ therefore requires comparable energy injection into electrons and neutrinos, $B_\nu\simeq B_{ee}\simeq1/2$. Gradual decay and energy exchange between the two sectors modify this estimate. Even so, the total neutrino branching fraction primarily sets $\neff$, while the relative contributions of the neutrino flavors affect neutron conversion. Mixed decays can therefore satisfy the radiation and abundance constraints at low reheating temperatures.

For lighter reheaters, the softer antineutrinos are closer to the threshold of $\bar\nu_e+p\to n+e^+$, which suppresses this reaction relative to $\nu_e+n\to p+e^-$. Unlike for heavier masses, the injection then shifts the neutron-proton balance toward protons, so reheating can end at still lower temperatures without excessive helium. The coldest compatible histories therefore favor a large $B_{\nu_e}$. Muon and tau neutrinos do not react with nucleons at these energies, so the flavor fractions change $\yp$ at nearly fixed $\neff$; oscillations turn part of the $\nu_\mu$ and $\nu_\tau$ into $\nu_e$ and can lower an excessive abundance into the measured range.

The gain from lighter reheaters is limited, because softer injection converts fewer neutrons at a given neutrino energy density. Later decays also leave less time before nuclei form. The helium and deuterium constraints together then set the minimum temperature: the still colder histories examined down to $\Trh\simeq0.25\,\MeV$ give $\yp\gtrsim0.36$. In the compatible examples near $\Trh\simeq0.4\,\MeV$, radiation domination begins around $T\simeq0.5\,\MeV$, with negligible reheater energy by $T\simeq0.2\,\MeV$. Reheating therefore changes weak freeze-out well before nuclear burning.

In the gap at intermediate temperatures in Fig.~\ref{fig:minima} (right), the softened electron neutrino and antineutrino spectra suppress proton-to-neutron conversion more strongly than the reverse reaction. Continued weak interactions during cooling then leave too few neutrons to reproduce the measured helium abundance. Hotter reheating reduces this spectral bias; at lower temperatures, the overall reaction rates fall enough for more neutrons to survive freeze-out. Deuterium underproduction in the central part of the gap strengthens the joint exclusion.

Finally, the full neutrino evolution does not reproduce the allowed region of Ref.~\cite{Hannestad:2004px} near $\Trh=1\,\MeV$, where $B_\nu\approx0.9$. These decays overproduce radiation, as Eq.~\eqref{eq:branch-balance} suggests, while energetic antineutrinos convert protons into neutrons. For example, $m_X=20\,\MeV$, $\Gamma_X=4\,\mathrm{s}^{-1}$, and $B_\nu=0.9$ give $\neff\simeq9.6$ and $\yp\simeq0.33$ (see SuM), compared with the calculated $3<\neff<5$ and adopted helium range $0.238\pm0.015$ of Ref.~\cite{Hannestad:2004px}.

\paragraph{Particle realizations.}
Mixed decays require $X$ to couple separately to electrons and to neutrinos, so that the rates of the two channels can be chosen independently~\cite{Babu:2019iml}. For $m_X\sim10\,\MeV$ and $\Trh\simeq0.4\,\MeV$, the required couplings are of order $10^{-11}$, too weak for laboratory searches~\cite{NA64:2021xzo,Pasquini:2015fjv}.
Stellar production nevertheless constrains them:
if they are as large today as during reheating, scalars from
supernova cores decay into $e^+e^-$ outside stripped progenitors,
depositing more energy than observations of SN 1998bw
allow~\cite{Candon:2025ypl}. Weakening them after reheating avoids this bound. An ultralight field that starts to oscillate, or a scalar that slowly rolls afterwards, can do so~\cite{Gan:2023wnp,Damour:2002mi}, for example through the mass mixing of $X$ with an electrophilic state, which LDMX could test~\cite{Berlin:2018bsc,Gaiser:2026kkx}. An explicit model of this type satisfies the joint BBN and CMB constraints near $\Trh\simeq0.4\,\MeV$ (see SuM). The reheater may also consist of non-topological solitons, which stars cannot form and whose free quanta can be too heavy for stellar emission~\cite{Kawasaki:2002hq,Kawasaki:2022hvx}; these might be compatible as well. Late reheating through electrons and neutrinos can thus arise in different realizations of $X$. The bounds of Fig.~\ref{fig:minima} apply to a real CP-even scalar with the decays and inverse reactions specified in the SuM, and other spins or couplings can shift them. Detailed model building is left to future work.

\begin{figure}[t!]
 \centering
 \includegraphics[width=\columnwidth]{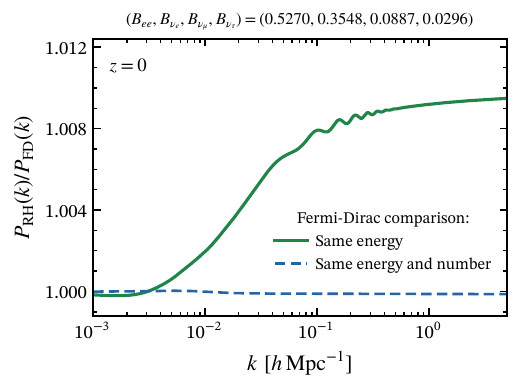}
 \caption{Ratios of linear matter power today, Eq.~\eqref{eq:active-power}, for $m_X=6\,\MeV$ and $\Trh=0.42\,\MeV$, with the decay fractions indicated above the frame. We assume normal neutrino mass ordering with lightest mass $m_1=0$. The reference Fermi-Dirac shapes match each state's relativistic energy density (green) or both number and energy densities (blue), at fixed physical neutrino masses and total matter density.}
 \label{fig:active-mass}
\end{figure}

{
\paragraph{Implications.}
Late reheating reshapes the relic neutrino populations and dilutes any species produced before it, so its consequences reach beyond $\neff$ and the light elements. We illustrate them with active and sterile neutrinos. Active neutrinos can carry nearly standard radiation density with fewer particles if their mean momenta are higher. Once they become non-relativistic, the smaller number lowers their mass density at the same physical masses~\cite{Oldengott:2019mass,Alvey:2021distribution}. The decay pattern in Fig.~\ref{fig:active-mass} particularly depletes the heaviest state. With $\neff\simeq2.89$, the late neutrino mass density then falls to about a third of its standard value for normal ordering with $m_1=0$. For an $11\,\MeV$ reheater with $B_{\nu_e}\simeq0.92B_\nu$, it reaches about a fifth.

We compute linear matter power with \textsc{Class}~\cite{Blas:2011rf,Lesgourgues:2011rh}, using the mass eigenstate distributions from \nudsmc and adiabatic primordial perturbations. At comoving wavenumber $k$ and redshift $z=0$, the ratio of reheating to Fermi-Dirac power is
\begin{equation}
 R_P(k)=\frac{P_{\rm RH}(k,z=0)}{P_{\rm FD}(k,z=0)}.
 \label{eq:active-power}
\end{equation}
The comparison fixes the physical masses, each state's relativistic energy density, and total matter density. The smaller neutrino mass density weakens the suppression of clustering, raising matter power by $\simeq1\%$ (Fig.~\ref{fig:active-mass}). Matching the number densities as well removes nearly all of this difference.

These changes affect cosmological neutrino mass measurements through galaxy clustering and lensing. Euclid and the Simons Observatory target sensitivities of a few $10^{-2}\,\mathrm{eV}$ in combination with other data, assuming thermal relics~\cite{Euclid:2024neutrinos,SimonsObservatory:2018koc}. Late reheating breaks this premise, so it changes the relation between those observables and physical masses. In the example with normal ordering, a mass sum of $0.059\,$eV gives the present mass density of thermal neutrinos with $\sum_im_i\simeq0.019\,$eV. This would ease the tension between oscillation data and fits combining CMB and DESI measurements, which prefer sums below $0.06\,$eV~\cite{DESI:2025DR2}.

Another implication concerns the sector of sterile neutrinos $\nu_s$ with large couplings to the Standard Model. A $\nu_s$ with a mass of a few tens of eV and active mixing $|U_{\alpha4}|^2\sim10^{-3}$ to $10^{-2}$ is a target for laboratory searches. Here, $U_{\alpha4}$ is the mixing amplitude between active flavor $\alpha$ and the predominantly sterile mass eigenstate $\nu_4$. KATRIN directly probes sterile neutrinos through their mixing with electron neutrinos~\cite{KATRIN:2025lph}, while nuSCOPE, SHiP, and forward LHC neutrino detectors search for oscillations~\cite{Blanchet:2026jwx,Choi:2025vby,Anchordoqui:2021ghd}. However, these states would exceed the observed dark matter density if thermalized. Even a plasma heated instantaneously to $5\,\MeV$ can produce too much hot dark matter for muon mixing within the projected experimental reach~\cite{Gelmini:2004ah,Peters:2023asu}.

Reheating through decays into electrons and neutrinos suppresses sterile neutrino production further. For a state mixed with $\nu_\mu$, choosing $B_{\nu_\mu}=0$ eliminates its direct production in $X$ decays. Near ${\Trh=1\,\MeV}$, its abundance can then fall by a factor of $\simeq40$ relative to instantaneous reheating to $5\,\MeV$, leaving percent-level suppression of matter clustering. Such abundances occur at mixings within the projected sensitivity of nuSCOPE and of SHiP with a proposed far detector~\cite{Blanchet:2026jwx,Choi:2025vby}.
}

\paragraph{Conclusions.}
Decays of a light particle into electrons and neutrinos allow reheating temperatures around $0.4\,\MeV$, compatible with joint BBN and CMB constraints. Directly produced neutrinos make up for the deficit left by heating the EM plasma alone, and the branching fractions into each flavor change the neutron-to-proton ratio even with oscillations. Heavier particles inject more energetic neutrinos, which convert protons into neutrons and strengthen the bound; above their threshold, charged pions raise it for EM reheating as well.

Reheating this late can also strongly alter the number and momentum distributions of active neutrinos while keeping $\neff$ nearly standard. At unchanged physical masses, fewer and more energetic neutrinos then yield a smaller present mass density, by up to a factor of five for normal ordering. Cosmological fits of neutrino masses constrain this density and therefore depend on the flavor fractions of the decays. Masses at the minimum allowed by oscillation experiments could then appear smaller in these fits, easing the tension between the two. Late reheating can also make sterile neutrinos with mixings large enough for laboratory searches compatible with limits on hot dark matter.

\section*{Acknowledgements}
We thank Miguel Escudero for his involvement in the early stages of this project, Kazunori Kohri for useful discussions about late reheating, Andrea Caputo for conversations about supernova bounds on light particles, and Matei Climescu for carefully reading the manuscript.
KA was supported by JSPS KAKENHI Grant Number 24KJ0060.
MO received support from the European Union's Horizon Europe research and
innovation programme under the Marie Sk{\l}odowska-Curie grant agreement No.~101204216.
OI received funding from the European Union's Horizon Europe research and innovation programme under the Marie Sk{\l}odowska-Curie grant agreement No.~101261809.

The authors used Astra 6 and Claude Opus 5.5 for assistance with the optimization of the \textsc{C++} code and the cross-testing of model realizations of late reheating scenarios with light particles.
The authors take full responsibility for the results.
\bibliography{main}

\onecolumngrid
\clearpage
\begin{center}{\bf\large Supplemental Material}\end{center}
\setcounter{equation}{0}
\setcounter{figure}{0}
\setcounter{table}{0}
\setcounter{section}{0}
\setcounter{secnumdepth}{1}
\renewcommand{\theequation}{S\arabic{equation}}
\renewcommand{\thefigure}{S\arabic{figure}}
\renewcommand{\thetable}{S\arabic{table}}
\renewcommand{\thesection}{S\arabic{section}}
\renewcommand{\theHequation}{S\arabic{equation}}
\renewcommand{\theHfigure}{S\arabic{figure}}
\renewcommand{\theHtable}{S\arabic{table}}
\renewcommand{\theHsection}{S\arabic{section}}

This Supplemental Material is organized as follows. Section~\ref{sec:sm-transport} describes reheating evolution and the joint cosmological constraints. Section~\ref{sec:sm-comparison} compares neutrino evolution and nuclear abundances with other kinetic calculations and tests the numerical resolution of \nudsmc. Section~\ref{sec:sm-heavy-alternatives} examines dark radiation and enhanced neutrino scattering. Sections~\ref{sec:sm-scalar} and~\ref{sec:sm-pions} discuss particle models and secondary pion production, respectively. Section~\ref{sec:sm-relics} covers active neutrino mass signals, sterile neutrino production, and dilution of axion-like condensates.

\section{Reheating evolution and cosmological constraints}
\label{sec:sm-transport}
The bounds in Fig.~\ref{fig:minima} rest on the reheater population, the \nudsmc calculation, and the likelihood specified below.
\subsection{Late reheating in \texorpdfstring{\textmd{\nudsmc}}{nuDSMC}}
The initial scalar abundance and its in-medium reaction rates determine when energy reaches the plasma. At a common photon and neutrino temperature of $20\,\MeV$, we take one bosonic degree of freedom at rest with ${\rho_X/\rho_{\rm rad}=7\times10^5\,m_X/\MeV}$, where $\rho_{\rm rad}$ is the equilibrium radiation density.
We assume a non-thermal origin for this spatially homogeneous
population, which dominates the cosmic energy budget before decaying.
As in Refs.~\cite{Kawasaki:1999na,Kawasaki:2000en,Hannestad:2004px}, we neglect density perturbations of the reheater. They grow while $X$ dominates~\cite{Erickcek:2011us} and make the baryon-to-photon ratio vary in space. Such variations over distances above the neutron diffusion length raise the averaged $\dh$~\cite{Barrow:2018yyg,Bagherian:2025puf} by an amount set by the primordial power on small scales and by the evolution of the reheater before $T=20\,\MeV$.

We initially treat the bath as equilibrated while evolving scalar decays, inverse reactions, and expansion. Momentum-resolved neutrino evolution begins when the thermal neutron-proton relaxation rate falls to $2H$, or when $T$ reaches $5\,\MeV$, whichever occurs first. Here, $H$ includes the scalar energy density. The scalar populations, elapsed time, and neutron fraction are carried continuously into this evolution.

We use charge-symmetric spectra and average rapid oscillation phases~\cite{Akita:2026xwu}. Standard Model collision terms include interference between flavor amplitudes; coherence within neutrino pairs from $X$ decays and refraction from coherent neutrino-antineutrino density differences are omitted. Unless specified otherwise, the vacuum mixing matrix has normal mass ordering, CP phase $\delta_{\rm CP}=0$, mixing angles $\theta_{ij}$, and mass-squared differences $\Delta m^2_{ij}=m_i^2-m_j^2$ given by
\begin{equation}
 (\sin^2\theta_{12},\sin^2\theta_{23},\sin^2\theta_{13})=(0.297,0.425,0.0215),\qquad
 (\Delta m^2_{21},\Delta m^2_{31})=(7.37\times10^{-5},2.56\times10^{-3})\,\mathrm{eV}^2.
 \label{eq:sm-mixing}
\end{equation}

The thermal electron mass can suppress or close $X\to e^+e^-$. For the electron Yukawa coupling $-y_eX\bar ee$, the vacuum and thermal electron masses, $m_{e,0}$ and $m_{e,\rm eff}(T)$, modify the partial width at plasma temperature $T$ as
\begin{equation}
 \frac{\Gamma_{ee}^{\rm med}(T)}{\Gamma_{ee}^{\rm vac}}
 =\left[\frac{1-4m_{e,\rm eff}^2(T)/m_X^2}{1-4m_{e,0}^2/m_X^2}\right]^{3/2}
 \Theta(m_X-2m_{e,\rm eff}(T)),
 \label{eq:sm-ee-medium}
\end{equation}
where $\Gamma_{ee}^{\rm med}$ is the in-medium width and $\Theta$ is the Heaviside function; the bracket is evaluated only above threshold. The vacuum width fixes $y_e$. Pauli blocking is applied separately, and the inverse reaction uses the same matrix element and kinematics. These in-medium factors change the relative rates, while the vacuum branching fractions stay fixed.

\subsection{Nuclear abundances and cosmological constraints}
\label{sec:sm-bbn}
The calculation of nuclear abundances follows neutron-proton conversion and nuclear reactions through the evolving neutrino spectra and entropy release. We use \textsc{Primat} AC2024 nuclear reaction rates, finite nucleon mass corrections to the weak rates, quantum electrodynamic (QED) thermodynamics, the neutron fraction carried from the initial evolution, and secondary pion production~\cite{Akita:2026xwu}. With subscript $f$ denoting a final epoch after entropy release, the baryon-to-photon ratio is
\begin{equation}
 \eta_B(t)=273.9\,\omega_b\,10^{-10}
 \left[\frac{a_fT_f}{a(t)T(t)}\right]^3.
 \label{eq:eta-history}
\end{equation}
The changing baryon density affects nuclear reactions and pion absorption. Neutrino capture also depends on the neutrino momentum distributions and reaction thresholds. For incident neutrino energy $E$, neutron-proton mass difference $\Delta=m_n-m_p$, and electron mass $m_e$, the capture cross sections on stationary nucleons, without final-state blocking, obey
\begin{equation}
 \frac{\sigma_{\bar\nu_ep\to ne^+}}{\sigma_{\nu_en\to pe^-}}
 \simeq\frac{(E-\Delta)p_-}{(E+\Delta)p_+},\qquad
 p_\pm=\sqrt{(E\pm\Delta)^2-m_e^2},\qquad E>\Delta+m_e,
 \label{eq:capture-ratio}
\end{equation}
where $p_+$ and $p_-$ are the outgoing electron and positron momenta~\cite{Pitrou:2018cgg}. Antineutrino-induced neutron production vanishes below threshold. For injection at a few MeV, lowering $E$ at fixed $\rho_{\nu_e}$ also reduces the capture rate per neutron, $\lambda_{\nu_en}\propto\rho_{\nu_e}(E+\Delta)p_+/E$.

The nuclear reaction network uses weak rates obtained by integrating the corrected cross sections over the evolving neutrino distributions. These spectra are far from thermal while the reheater decays. In the example of Fig.~\ref{fig:qke-spectra}, between $T=4$ and $1\,\MeV$, $16$ to $33\%$ of the electron neutrinos would have to change momentum to reproduce the Fermi-Dirac distribution with the same number and energy densities, whose temperature and chemical potential are both free. Computed with the cross sections of Eq.~\eqref{eq:capture-ratio}, the rates of $\nu_e+n\to p+e^-$ and $\bar\nu_e+p\to n+e^+$ in this thermal approximation deviate from the actual values by up to $11\%$ and $20\%$, respectively. After the decays, the mismatched fraction falls to about $6\%$. Neutrino number and energy densities therefore do not determine neutron-proton conversion during reheating.

The joint constraint uses the correlated CMB-SPA likelihood with independently varied radiation density and helium~\cite{SPT-3G:2025bzu}, combining Planck, ACT, and SPT temperature, polarization, and lensing data without baryon acoustic oscillations (BAO). We denote the primordial helium-4 mass fraction used in recombination by $Y_{\rm CMB}$. In the convention of Eq.~\eqref{eq:Neff}, $\neff=N_{\rm ur}+1.0132$: $N_{\rm ur}$ is the effective contribution of ultrarelativistic species, and $1.0132$ accounts for the fixed $0.06\,\mathrm{eV}$ state while relativistic. Flat priors in $(\omega_b,\neff,Y_{\rm CMB})$ make the posterior of Ref.~\cite{SPT-3G:2025bzu} proportional to the likelihood marginalized over the remaining cosmological and nuisance parameters. We therefore evaluate it with a weighted Gaussian kernel density estimate of its Markov chains. In the ordered basis $\boldsymbol\theta=(\omega_b,\neff,Y_{\rm CMB})$, their mean $\overline{\boldsymbol\theta}$, standard deviations $\boldsymbol\sigma$, correlation matrix $R$, and covariance $C_{\rm CMB}$ are
\begin{equation}
 \begin{aligned}
 \overline{\boldsymbol\theta}&\simeq(0.02219,2.962,0.2329),\qquad
 \boldsymbol\sigma\simeq(0.000143,0.227,0.0132),\\
 R&\simeq\begin{pmatrix}1&0.371&0.126\\0.371&1&-0.782\\0.126&-0.782&1\end{pmatrix},
 \qquad C_{{\rm CMB},ij}=\sigma_iR_{ij}\sigma_j.
 \end{aligned}
 \label{eq:cmb-covariance}
\end{equation}
We restrict evaluation to the original prior ranges, $2.0132\leq\neff\leq6.0132$, $0.005\leq\omega_b\leq0.1$, and $0.1\leq Y_{\rm CMB}\leq0.5$. This three-parameter likelihood treats the massive neutrinos as thermal. Section~\ref{sec:sm-relics} tests this assumption for the populations left by late reheating.

The same nuclear uncertainty in helium enters the CMB and abundance likelihoods. Using hydrogen and helium atomic masses $m_{\rm H}$ and $m_{\rm He}$~\cite{Fields:2019pfx}, we convert the abundance defined by number densities $\yp$ to the recombination mass fraction through
\begin{equation}
 Y_{\rm CMB}=\mathcal Y(\yp)=\frac{(m_{\rm He}/4)\yp}{m_{\rm H}(1-\yp)+(m_{\rm He}/4)\yp}.
 \label{eq:helium-conversion}
\end{equation}
We describe nuclear uncertainties by shifts $\boldsymbol\delta=(\delta_Y,\delta_D)$ in $(\yp,\Dfive)$ with normalized zero-mean Gaussian density $\mathcal N$ and covariance $C_{\rm nuc}$. The joint likelihood is
\begin{equation}
 \mathcal L_{\rm joint}=\int d^2\boldsymbol\delta\,
 \mathcal N(\boldsymbol\delta;0,C_{\rm nuc})\,
 \mathcal L_{\rm CMB}(\omega_b,\neff,\mathcal Y(\yp+\delta_Y))\,
 \mathcal L_{\rm abund}(\yp+\delta_Y,\Dfive+\delta_D).
 \label{eq:joint-covariance}
\end{equation}
The abundance likelihood $\mathcal L_{\rm abund}$ uses Eq.~\eqref{eq:bbn_data}, whose helium measurement is expressed as $\yp=4n_{{}^4\mathrm{He}}/n_b$~\cite{Yeh:2026lbt}. We use nuclear uncertainties evaluated as functions of baryon density from 1000 variations of \textsc{Primat} reaction rates and neutron lifetime $\tau_n$, centered on $878.4\,\mathrm{s}$~\cite{PRIMAT:2026errors}, and neglect their correlation between helium and deuterium. At $\omega_b=0.0223$, the standard deviations of $\yp$ and $\Dfive$ are $1.1\times10^{-4}$ and $0.026$. A full marginalization over nuclear reaction rates, including their correlations with cosmological parameters, affects abundance constraints~\cite{Akharman:2026izy}. We approximate it by the Gaussian shifts in Eq.~\eqref{eq:joint-covariance} and use the standard BBN uncertainties also for non-thermal reheating.

We profile the common baryon density and decay fractions using
\begin{equation}
 \chi^2=-2\ln\mathcal L_{\rm joint}+\mathrm{constant}.
 \label{eq:criterion}
\end{equation}
The nuclear covariance normalization is included. For the quoted $\chi^2$, we normalize the CMB term at $\overline{\boldsymbol\theta}$ and the abundance covariance at $\omega_b=0.0223$; the additive constant cancels in likelihood ratios.
The minimum joint $\chi^2$ is $9.40$ over the reheating masses,
temperatures, and decay fractions considered, with $\omega_b$
profiled. The separate CMB and BBN regions of Fig.~\ref{fig:minima} (right) use their own likelihoods and best-fit references in the same way. The compatible histories at the lowest $\Trh$ have
$\neff\simeq2.89$ to $3.21$
(Table~\ref{tab:examples}), close to the standard value, so the
bounds on $\Trh$ do not rely on a reduced radiation density.

The asymptotic thresholds in Eqs.~\eqref{eq:minimum} and~\eqref{eq:joint-region} have approximate coverage near parameter boundaries. For electron-only decays below $m_X=100\,\MeV$, $\chi^2$ still decreases at the highest reheating temperature we calculate, $8$ to $14\,\MeV$ depending on $m_X$, and we use this value as $\chi^2_{\min}$. A smaller $\chi^2$ above this range would lower $\chi^2_{\min}$ and raise $q_{m_X}$ of Eq.~\eqref{eq:minimum} at every $\Trh$. The condition $q_{m_X}\leq4$ would then first hold at a larger $\Trh$, so the plotted $\Trhlow$ is conservative. At $m_X\geq100\,\MeV$, standard BBN, which corresponds to large $\Trh$, has a smaller $\chi^2$ than all calculated histories and serves as $\chi^2_{\min}$.

Table~\ref{tab:examples} gives four compatible decay patterns used in the studies of neutrino masses and other relics below. The standard errors from ten independent Monte Carlo realizations of the $\Trh=0.42\,\MeV$ example are $0.003$ in $\neff$ and $4\times10^{-5}$ in $\yp$.

The flavor families of Fig.~\ref{fig:minima} specify the neutrino compositions $(B_{\nu_e},B_{\nu_\mu},B_{\nu_\tau})/B_\nu$ over which we profile, and each is calculated for several values of $B_{ee}$. The basic set contains the quarter steps $(i,j,4-i-j)/4$ with nonnegative integers $i+j\leq4$, equal fractions, and a continuous ratio $B_{\nu_\mu}:B_{\nu_\tau}$ at $B_{\nu_e}=0$. Where the reheating temperatures are lowest, for $5\leq m_X/\MeV\leq11.5$ and $0.35\leq\Trh/\MeV\leq1.1$, we add compositions with $\xi_e\equiv B_{\nu_e}/B_\nu$ between $0.7$ and $0.92$. Between them, the observables are interpolated linearly.

Near the lowest reheating temperatures, the compatible histories have $B_{ee}\simeq0.49$ to $0.54$, as the energy balance of Eq.~\eqref{eq:branch-balance} suggests. Their preferred $\xi_e$ decreases as $\Trh$ rises, while the split between muon and tau neutrinos is weakly constrained.

\begin{table}[t!]
\centering
\setlength{\tabcolsep}{5pt}
\begin{tabular}{ccccccccc}
$m_X/\MeV$ & $\Trh/\MeV$ & $(B_{ee},B_{\nu_e},B_{\nu_\mu},B_{\nu_\tau})$ & $\omega_b$ & $\neff$ & $\yp$ & $\Dfive$ & $\chi^2$ & $n$\\\hline
6 & 0.425 & $(0.5270,0.3548,0.0887,0.0296)$ & 0.02215 & 2.891 & 0.24583 & 2.439 & 11.18 & 10\\
12.5 & 1.050 & $(0.4906,0.2547,0,0.2547)$ & 0.02231 & 3.213 & 0.24583 & 2.492 & 13.43 & 10\\
11.25 & 1.050 & $(0.5218,0,0.4782,0)$ & 0.02218 & 2.959 & 0.24518 & 2.448 & 10.84 & 10\\
200 & 5.400 & $(1,0,0,0)$ & 0.02220 & 2.943 & 0.24809 & 2.455 & 12.97 & 1
\end{tabular}
\caption{Reheater decay patterns and their nuclear abundances, radiation density, and likelihood statistic, Eq.~\eqref{eq:criterion}. All rows satisfy the joint constraint, with baryon density profiled; the active neutrino, sterile neutrino, and condensate calculations use the first, second, and third and fourth rows, respectively. The number $n$ counts independent realizations, whose observables are averaged. Fractions and observables are rounded.}
\label{tab:examples}\label{tab:near-standard}\label{tab:relic-histories}
\end{table}

\section{Comparisons with other calculations}
\label{sec:sm-comparison}
The bounds of Fig.~\ref{fig:minima} are calculated with \nudsmc. To test its neutrino spectra and radiation densities, we solve selected histories with quasi-classical Boltzmann equations (QCBE) and quantum kinetic equations (QKE), described in detail in Ref.~\cite{Akita:2026xwu}. We also refine the numerical resolution of \nudsmc for the history with the lowest $\Trh$ in Table~\ref{tab:examples}.

For electromagnetic reheating, the neutrino spectra agree with QCBE in the example examined. For mixed decays, \nudsmc gives radiation densities close to QCBE when oscillations are omitted and to QKE when they are included. Without flavor conversion, the quantum kinetic equations reduce exactly to QCBE, and \nudsmc agrees with both (Table~\ref{tab:sm-neff-comparison}). We also revisit a low-temperature example allowed by the earlier joint cosmological analysis~\cite{Hannestad:2004px}. In the mixed cases of this section, the criterion of Sec.~\ref{sec:sm-transport} starts momentum-resolved neutrino evolution at $T=5\,\MeV$.

\begin{table}[t!]
 \centering
 \begin{tabular}{cccccccc}
 \hline
 $m_X/\MeV$ & $(B_{\nu_e},B_{\nu_\mu},B_{\nu_\tau})/B_\nu$ & Inverse reactions & Oscillations & \nudsmc & QCBE & QKE-ATAO & Full QKE\\\hline
 $12.5$ & $(0.75,0.25,0)$ & off & off & $2.75$ & $2.75$ & -- & $2.76$\\
 $12.5$ & $(0.75,0.25,0)$ & off & on & $3.0200\pm0.0006$ & -- & $3.0207$ & $3.005$\\
 $12.5$ & $(0.75,0.25,0)$ & on & off & $2.747\pm0.008$ & $2.754$ & -- & --\\
 $10$ & $(0,0.25,0.75)$ & on & off & $3.126\pm0.006$ & $3.128$ & -- & --\\\hline
 \end{tabular}
 \caption{Radiation density $\neff$, Eq.~\eqref{eq:Neff}, after mixed decays of $X$ into electrons and neutrinos with $B_{ee}=B_\nu=0.5$ at $\Trh=1\,\MeV$. It is calculated with \nudsmc, QCBE, and QKE, solved in the ATAO approximation (QKE-ATAO) and in full; dashes mark calculations not performed. The \nudsmc uncertainties are standard errors of the means of three realizations.}
 \label{tab:sm-neff-comparison}
\end{table}

Electromagnetic reheating gives close agreement between \nudsmc and QCBE with oscillations disabled~\cite{Akita:2024nam,Akita:2026xwu}. Both evolve separate flavor occupation numbers $f_{\nu_\alpha}(p,t)$ and $f_{\bar\nu_\alpha}(p,t)$, so we compare their spectra. At $\Trh\simeq2.65\,\MeV$ and $T\simeq1\,\MeV$, the electron neutrino energy moments $\int dy\,y^3f_{\nu_e}(y)$, with $y=p/T$, differ by about $0.15\%$.

Mixed decays give close radiation densities in the comparisons of \nudsmc with QCBE and QKE. We use $m_X=12.5\,\MeV$, $\Trh=1\,\MeV$, and $(B_{ee},B_{\nu_e},B_{\nu_\mu},B_{\nu_\tau})=(0.5,0.375,0.125,0)$, omitting inverse decays. Without oscillations, \nudsmc and QCBE give the same $\neff$. Flavor conversion changes energy transfer between the EM plasma and neutrinos, raising the final $\neff$ by about $0.27$ in both \nudsmc and QKE. Thus, the change caused by oscillations is substantially larger than the difference between the kinetic calculations. With and without flavor conversion, their values of $\neff$ for this history agree within $0.016$ (Table~\ref{tab:sm-neff-comparison}).

For the comparison with oscillations, \nudsmc uses collision interference, normal ordering, $\delta_{\rm CP}=0$, $(\sin^2\theta_{12},\sin^2\theta_{23},\sin^2\theta_{13})=(0.307,0.561,0.02195)$, and $(\Delta m^2_{21},\Delta m^2_{31})=(7.49\times10^{-5},2.534\times10^{-3})\,\mathrm{eV}^2$. Three realizations determine its mean $\neff$ to $0.0006$ (Table~\ref{tab:sm-neff-comparison}). Changing particle numbers and collision grouping shifts it by less than $0.002$.

\begin{figure}[t!]
 \centering
 \includegraphics[width=0.49\textwidth]{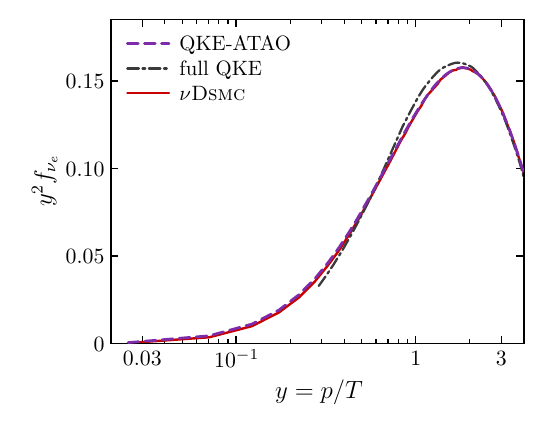}\hfill
 \includegraphics[width=0.49\textwidth]{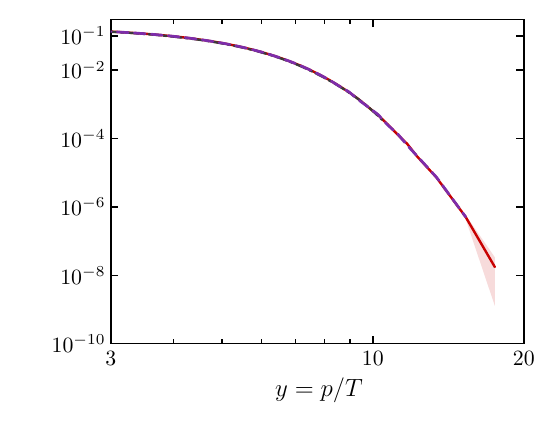}
 \caption{Electron neutrino spectra from mixed scalar decays without inverse reactions, comparing \nudsmc with QKE, both full and in the ATAO approximation~\cite{Froustey:2020mcq}. The parameters are $m_X=12.5\,\MeV$, $\Trh=1\,\MeV$, and $(B_{ee},B_{\nu_e},B_{\nu_\mu},B_{\nu_\tau})=(0.5,0.375,0.125,0)$. We plot $y^2f_{\nu_e}$ at $T=0.024\,\MeV$, proportional to the number density per momentum interval in Eq.~\eqref{eq:sm-qke-moments}. Here, $y=p/T$ uses the neutrino momentum in the cosmological rest frame, and $f_{\nu_e}=f_{\bar\nu_e}$ is the occupation per helicity. Left: the bulk distribution on a linear vertical scale. Right: the high-momentum tail on a logarithmic vertical scale. The \nudsmc result is the mean of three realizations, with a band showing its standard error. The full QKE result is not yet converged numerically; it uses a momentum grid with spacing $\Delta y\simeq0.29$ and, above $y=4$, is averaged over groups of three grid intervals.}
 \label{fig:qke-spectra}
\end{figure}

Spectra of $\nu_e$, $\nu_\mu$, and $\nu_\tau$ test whether the agreement in total radiation density extends to the distribution of neutrino energy among momenta and flavors. We perform an independent three-flavor calculation using the adiabatic transfer of averaged oscillations (ATAO) approximation~\cite{Froustey:2020mcq}. Its results, denoted QKE-ATAO, enter only the cross-check of mixed decays with flavor conversion in Tables~\ref{tab:sm-neff-comparison} and~\ref{tab:qke-spectra} and Fig.~\ref{fig:qke-spectra}. This approach evolves the occupations of the instantaneous matter eigenstates and averages out their rapidly oscillating coherences. We reconstruct the flavor density matrices to evaluate the full matrix collision integrals and the Pauli-blocked decay source, then project their derivatives back. The Hamiltonian contains vacuum mixing, with the parameters specified above, and the electron-positron matter potential. Neutrino mean-field potentials are omitted; neutrino-neutrino collisions are included.

The QKE-ATAO calculation uses the mixed decay fractions specified above, $\tau_X\simeq0.494\,\mathrm{s}$, and $\rho_X/\rho_{\rm rad}=8.75\times10^6$ at $20\,\MeV$. It evolves the background with QCBE from $20$ to $5\,\MeV$, where the neutrino distributions start in thermal equilibrium. The plasma includes finite-mass electrons and finite-temperature QED corrections, but no muons or pions. The partial width for $X\to e^+e^-$ has its vacuum value; \nudsmc applies the in-medium factor of Eq.~\eqref{eq:sm-ee-medium}, whose removal shifts its $\neff$ by less than $0.002$. Inverse decays are omitted in both calculations.

The ATAO momentum grid is uniform in $y_{\rm cm}=p/T_{\rm cm}$, where $T_{\rm cm}\propto a^{-1}$, with spacing $\Delta y_{\rm cm}\simeq1$ from $y_{\rm cm}=0.01$; it extends to $y_{\rm cm}=400$ to cover the decay line as it moves to higher $y_{\rm cm}$. The kinetic equations are integrated with relative and absolute tolerances of $10^{-5}$ and $10^{-6}$ until $T=0.024\,\MeV$, after scalar depletion.

To compare spectra at this temperature, we use $y=p/T$ and the occupation per helicity $f_\alpha=f_{\nu_\alpha}=f_{\bar\nu_\alpha}$. The combined neutrino and antineutrino number and energy densities obey
\begin{equation}
 n_\alpha=\frac{T^3}{\pi^2}\int dy\,y^2 f_\alpha(y),\qquad
 \rho_\alpha=\frac{T^4}{\pi^2}\int dy\,y^3 f_\alpha(y).
 \label{eq:sm-qke-moments}
\end{equation}
We evaluate $\neff$ from these physical densities using Eq.~\eqref{eq:Neff}, with $\rho_\gamma=\pi^2T^4/15$ and $\rho_{\rm DR}=0$. Coarsening the ATAO momentum grid changes the resulting $\neff$ by $0.003$, so the two calculations match within about $0.1\%$ (Table~\ref{tab:sm-neff-comparison}).

For the same history, the full QKE yield $\neff$ about $0.015$ below the QKE-ATAO and \nudsmc values, and an electron neutrino spectrum that agrees with theirs within a few percent (Fig.~\ref{fig:qke-spectra}). With oscillations turned off, these equations reduce exactly to QCBE, yet they give a slightly larger $\neff$ (Table~\ref{tab:sm-neff-comparison}). Above $y\approx5$, the full QKE occupation also varies by $10$ to $20\%$ between neighboring momentum nodes. These features indicate that this calculation has not yet converged numerically.

The electron neutrino spectra agree closely across the momenta carrying most of the particles (Fig.~\ref{fig:qke-spectra}). We compare the QKE-ATAO and \nudsmc distributions on the \nudsmc momentum bins with their absolute normalizations, and denote by $\mathcal N_{\alpha i}=n_{\alpha i}/T^3$ the number density in bin $i$. The spectral difference without cancellation between momenta is
\begin{equation}
 D_\alpha=\frac{\sum_i y_i\left|\mathcal N_{\alpha i}^{\rm ATAO}
 -\overline{\mathcal N}_{\alpha i}^{\rm DSMC}\right|}
 {\overline{\rho}_\alpha^{\rm DSMC}/T^4},
 \label{eq:sm-qke-spectral-difference}
\end{equation}
where $y_i$ is the center of bin $i$ and bars mark the ensemble mean. This measure weights the absolute differences by momentum and includes the Monte Carlo fluctuations. In Table~\ref{tab:qke-spectra}, which also lists $D_\alpha$, the flavor energy densities agree within the statistical errors of \nudsmc. By contrast, ATAO contains slightly more neutrinos at $y\lesssim1$, where its momentum grid resolves the decay source coarsely, which raises its number densities by about $0.2\%$.

\begin{table}[t!]
 \centering
 \begin{tabular}{lrrr}
 \hline
 Flavor & $(n_\alpha^{\rm ATAO}/\overline n_\alpha^{\rm DSMC}-1)\,[\%]$ & $(\rho_\alpha^{\rm ATAO}/\overline\rho_\alpha^{\rm DSMC}-1)\,[\%]$ & $D_\alpha\,[\%]$\\\hline
 $e$ & $+0.237$ & $+0.0576$ & $0.215$\\
 $\mu$ & $+0.179$ & $-0.00002$ & $0.220$\\
 $\tau$ & $+0.156$ & $-0.0102$ & $0.216$\\\hline
 \end{tabular}
 \caption{Flavor number densities, energy densities, and spectral differences between QKE-ATAO and \nudsmc for the mixed decays in Fig.~\ref{fig:qke-spectra}, at $T=0.024\,\MeV$. The moments include neutrinos and antineutrinos and follow Eq.~\eqref{eq:sm-qke-moments}; $D_\alpha$ is defined in Eq.~\eqref{eq:sm-qke-spectral-difference}.}
 \label{tab:qke-spectra}
\end{table}

With inverse reactions included, as in the reheating bounds, \nudsmc and QCBE also agree for mixed decays with $B_{ee}=B_\nu=0.5$ at $\Trh=1\,\MeV$, without oscillations. Most injected neutrinos are electron neutrinos for $m_X=12.5\,\MeV$ and tau neutrinos for $m_X=10\,\MeV$. In both cases, the two calculations differ by less than the standard error of the \nudsmc mean (Table~\ref{tab:sm-neff-comparison}).

\begin{table}[t!]
 \centering
 \begin{tabular}{lcc}
 \hline
 Refinement & $\Delta\yp$ & $\Delta\neff$\\\hline
 Subdivisions of the decay line $\times2$ & $(1\pm8)\times10^{-5}$ & $0.004\pm0.005$\\
 Spatial cells $\times2$ & $(-5\pm12)\times10^{-5}$ & $0.008\pm0.005$\\
 Time step $\times1/2$ & $(3\pm8)\times10^{-5}$ & $0.012\pm0.005$\\\hline
 \end{tabular}
 \caption{Response of the average helium abundance $\yp$ and radiation density $\neff$, Eq.~\eqref{eq:Neff}, to a finer discretization in \nudsmc, for $m_X=6\,\MeV$ and $\Trh=0.42\,\MeV$. Each case averages five realizations of about $10^6$ initial neutrinos, and the uncertainties are statistical.}
 \label{tab:sm-resolution}
\end{table}

The highly occupied zero-momentum scalar mode and its narrow decay line are treated separately from resolved moving particles~\cite{Akita:2026xwu}. Local occupations set Pauli blocking and Bose enhancement in decays and inverse reactions, while expansion redshifts daughters out of the line. The rate of energy transfer through the decay line is checked against the analytic result for one decay channel in an expanding background.

Finer numerical resolution changes the predictions a little (Table~\ref{tab:sm-resolution}). For $m_X=6\,\MeV$ and $\Trh=0.42\,\MeV$, we double the subdivisions of the decay line or the number of spatial cells, or halve the time step. The shifts of $\yp$ stay within their statistical errors. Halving the time step changes $\neff$ most and lowers $\chi^2$ by $0.2$.

\subsection{Comparison with the previous late reheating studies}
For purely electromagnetic reheating, a detailed comparison with previous calculations is presented in Ref.~\cite{Akita:2026xwu}. There, we find good agreement with independent kinetic codes for both $\neff$ and nuclear abundances.

For mixed decays, we revisit the low-temperature region of Ref.~\cite{Hannestad:2004px}. We choose $m_X=20\,\MeV$, $\Gamma_X=4\,\mathrm{s}^{-1}$, and $B_\nu=0.9$, equally divided among flavors, inside the joint 95\% CMB, large-scale structure, and BBN region of its Fig.~7. These coordinates correspond to $\Trh\simeq1.41\,\MeV$. Starting with $\rho_X/\rho_{\rm rad}=3.5\times10^6$ at $20\,\MeV$, one realization with approximately one million initial neutrinos gives $\yp\simeq0.327$ at $\omega_b=0.0224$. This far exceeds even the loose helium range $0.238\pm0.015$ used in Ref.~\cite{Hannestad:2004px}. That range was deliberately broadened to encompass different determinations from metal-poor extragalactic H\,{\sc ii} regions~\cite{Fields:1998hi,Izotov:1998helium}. $\neff$ is much higher as well: $9.56$, compared with $3<\neff<5$ in Fig.~3 of that work.

The radiation excess also appears in three other cases with equal flavor fractions at $B_\nu=0.9$: $m_X=10$, $20$, and $40\,\MeV$ with $\Trh=1\,\MeV$ give $8.1\lesssim\neff\lesssim19.3$. These three calculations use $\rho_X/\rho_{\rm rad}=7\times10^5\,m_X/\MeV$ at $20\,\MeV$, as in Sec.~\ref{sec:sm-transport}, and start with approximately one million neutrinos each. The last two also overproduce helium throughout $0.020\leq\omega_b\leq0.023$. The discrepancy thus persists in all tested examples with equal flavor fractions.

\section{Dark radiation and enhanced neutrino scattering}
\label{sec:sm-heavy-alternatives}
Following the main text, we consider two modifications of electromagnetic late reheating: additional decays into dark radiation and enhanced neutrino self-interactions. Each could offset the neutrino depletion that strengthens the BBN bound. We use the initial conditions and pre-reheating evolution of Sec.~\ref{sec:sm-transport}.

First, we add decays into dark radiation, modeled as massless collisionless particles, at fixed total lifetime of $X$ and without direct neutrino injection or secondary hadrons. This increases the radiation density and expansion rate while leaving the active neutrinos depleted. The resulting helium excess prevents a substantial relaxation of the reheating bound. For example, a $250\,\MeV$ reheater decaying into photons and dark radiation at $\Trh=3\,\MeV$, with dark branching fraction $B_{\rm DR}\simeq0.18$, gives nearly standard $\neff$ and $\yp\simeq0.274$ after varying the baryon density. Lowering $B_{\rm DR}$ slightly from $0.18$ brings deuterium closer to its measured value, while helium remains excessive.

Second, we enhance elastic scattering among tau neutrinos through a scalar $\phi$ of mass $10\,\mathrm{GeV}$ with negligible abundance. Its interaction Lagrangian is $-\lambda_{\tau\tau}\phi\nu_\tau\nu_\tau/2+\mathrm{h.c.}$, where the coupling $\lambda_{\tau\tau}$ is dimensionless and h.c.\ denotes the Hermitian conjugate. We compare $\lambda_{\tau\tau}=0$, $0.2$, and $0.3$ in calculations that include $\phi$ exchange, its interference with Standard Model amplitudes, and all collisions and oscillation potentials of Sec.~\ref{sec:sm-transport}. By reshaping the neutrino spectra, this interaction alters the rates of the weak reactions with the EM plasma~\cite{Boyarsky:2021yoh}. For EM reheating with $m_X=1\,\mathrm{GeV}$ and $\Trh=4.5\,\MeV$, the added scattering lowers $\neff$ and raises $\yp$, so the neutrino depletion and the helium excess persist.

Separately, we test whether the same self-interactions can make mixed decays of a heavy $X$ compatible with the joint CMB and BBN constraints. By redistributing the momenta of the injected tau neutrinos, the added scattering can soften their high-energy tail, which drives neutron production. We consider $m_X=0.5$ and $1\,\mathrm{GeV}$, $\Trh=1$ and $1.5\,\MeV$, and $B_{ee}=0.2$, $0.3$, and $0.5$, with $B_{\nu_\tau}=1-B_{ee}$. Helium remains excessive at all three couplings: $\yp>0.31$ at each of the baryon densities $\omega_b=0.0218$, $0.0222$, and $0.0226$. Scattering through $\phi$ softens the high-energy tail only by heating the other neutrinos, so it leaves the mean neutrino momentum unchanged. Hard electron neutrinos and antineutrinos therefore persist, with comparable capture cross sections on neutrons and protons, respectively, so the more numerous protons undergo net conversion into neutrons~\cite{Hannestad:2004px}.

Any scattering of two neutrinos into two neutrinos, whatever its strength or flavor structure, conserves the total neutrino number and energy~\cite{Grohs:2020xxd}. Apart from redshift, lowering the mean neutrino momentum therefore requires production or annihilation of neutrinos, or heating of the EM plasma by neutrinos. We neglect number-changing reactions mediated by $\phi$, which arise only at higher order in $\lambda_{\tau\tau}\leq0.3$ and in $E_\nu/m_\phi\leq0.05$, where $E_\nu$ is the neutrino energy. We also omit the refractive potential of $\phi$, which leaves the mean neutrino momentum unchanged but can alter flavor conversion and hence the electron neutrino population.

\section{Model realizations}
\label{sec:sm-scalar}\label{sec:lab-opacity}
Separate scalar couplings to electrons and neutrinos allow their relative heating to vary. We describe this realization and the suppression of its couplings after reheating needed to satisfy stellar bounds. For a real CP-even field, the lepton interaction Lagrangian $\mathcal L_\ell$ contains electron Yukawa coupling $y_e$ and active Majorana neutrino couplings $\lambda_\alpha$, with vacuum partial widths~\cite{Babu:2019iml}
\begin{align}
 \mathcal L_\ell&=-y_eX\bar ee-\frac12\sum_\alpha(\lambda_\alpha X\nu_\alpha\nu_\alpha+\mathrm{h.c.}),
 \label{eq:scalar-lab-couplings}\\
 \Gamma_{ee}&=\frac{y_e^2m_X}{8\pi}\left(1-\frac{4m_e^2}{m_X^2}\right)^{3/2},\qquad
 \Gamma_{\nu_\alpha}=\frac{|\lambda_\alpha|^2m_X}{16\pi}.
 \label{eq:scalar-widths}
\end{align}
Each neutrino width includes charge-conjugate final states. We assume couplings diagonal in the charged lepton flavor basis. Their magnitudes are fixed by the lifetime and branching fractions through $\Gamma_c=B_c\Gamma_X$.

The same reheating mechanism applies to Dirac neutrinos when a vector couples to the active neutrino current and produces active neutrino-antineutrino pairs. By contrast, lepton-number-conserving scalar or pseudoscalar Yukawa interactions with Dirac neutrinos populate both active and sterile helicities~\cite{Babu:2019iml}; calculating reheating in those models requires evolving both populations. The reheater's spin, degrees of freedom, and reaction rates determine the quantitative temperature bound.

For a gauge-singlet $X$, the electroweak-invariant operators $X\bar L_eHe_R/\Lambda_e$ and $X(L_\alpha\!\cdot H)(L_\beta\!\cdot H)/\Lambda_\nu^2$ have independent coefficients. Here, $L_\alpha$ and $H$ are lepton and Higgs doublets, $e_R$ is the right-handed electron, and $\alpha,\beta=e,\mu,\tau$. The dot denotes the antisymmetric $SU(2)_L$ contraction; $\Lambda_e$ and $\Lambda_\nu$ are the energy scales suppressing the two effective operators. Heavy vectorlike charged leptons and Majorana singlets can generate these interactions. Their abundances and additional reactions are assumed negligible during reheating.

The required couplings, at most a few times $10^{-11}$, lie below laboratory sensitivity~\cite{NA64:2021xzo,Pasquini:2015fjv,Bjorken:1988as,Andreas:2012mt,Fan:2022eto}. Dense stellar matter nevertheless produces scalars efficiently through neutrino inverse decays, electron reactions, and plasmon processes~\cite{Akita:2022etk,Fiorillo:2022cdq,Carenza:2021pcm,Fiorillo:2025sln,Ferreira:2025qui,Hardy2025ScalarSN}. Observations of SN 1987A, supernova energy loss, and electron-pair fireballs constrain particles escaping the star~\cite{Akita:2023iwq,Blinov:2025aha,Chupp:1989kx,Caputo:2021rux,Hoof:2022xbe,Caputo2022LowEnergySN,Reguitti:2020psq,Kozyreva:2022wkp,Valerin:2022fxg,Candon:2025ypl,Soderberg:2008uh,Diamond:2023cto,Caputo:2025avc}. For the near-MeV reheating examples, decay lengths of $10^{11}$ to $10^{12}\,$cm allow the scalars to decay beyond stripped stars. The energy deposited outside the star is at least fifty times the SN 1998bw limit. Production through neutrino inverse decays alone exceeds this limit by a factor of ten or more.

A common dimensionless factor $u(t)$ can suppress the couplings after reheating and evade these stellar constraints. With subscript RH denoting their values during reheating, when $u=1$, we write
\begin{equation}
 y_e(t)=y_{e,\rm RH}u(t),\qquad \lambda_\alpha(t)=\lambda_{\alpha,\rm RH}u(t).
 \label{eq:evolving-couplings}
\end{equation}
The branching fractions stay fixed, while decay widths and stellar production rates scale as $|u|^2$. For example, an ultralight field $\varphi$ with $u=\varphi/\varphi_{\rm RH}$ can remain constant during reheating and subsequently oscillate about zero, with amplitude decreasing as $a^{-3/2}$~\cite{Gan:2023wnp}. Its mass and abundance must satisfy constraints from black-hole superradiance and cosmological structure formation~\cite{Stott:2018opm,Rogers:2020ltq,Dalal:2022rmp}; in particular, Ref.~\cite{Stott:2018opm} excludes $7\times10^{-20}<m_\varphi<10^{-16}\,$eV. A discrete symmetry relating field shifts to identical, unpopulated copies of the matter sector suppresses radiative corrections to the potential of $\varphi$ from the lepton couplings~\cite{Hook:2018jle,Gan:2023wnp,Delaunay:2025pho}.

A modulus that rolls down an exponential slope after reheating also reduces $u$, as a power of time, like a runaway dilaton~\cite{Damour:2002mi}. Its potential must stay shallow enough to hold it during the decays, despite terms induced by the lepton couplings, and its later contribution to the expansion must respect limits on early dark energy~\cite{Gomez-Valent:2022bku}. Since fireballs, SN 1987A gamma rays, and envelope heating all probe $e^+e^-$ pairs, weakening only the electron width can suffice. An evolving mass mixing with an electrophilic state achieves this while the neutrino couplings stay constant, provided the vacuum energy it generates falls as it shrinks. Ref.~\cite{Goertz:2024gzw} constructs a comparable time dependence for sterile neutrinos.

\begin{table}[t!]
 \centering
 \begin{tabular}{lc}
 \hline
 Quantity & Value\\\hline
 Mass of $X$, $m_X$ & $7.5\,\MeV$\\
 Coupling of $X$ to electron neutrinos & $2.5\times10^{-11}$\\
 Mass of $S$ & $35\,\MeV$\\
 Electron Yukawa coupling of $S$, $y_S$ & $1.5\times10^{-5}$\\
 Mass of the vectorlike lepton & $350\,\mathrm{GeV}$\\
 Mass of the triplet $\Delta$ & $1.5\,\mathrm{TeV}$\\
 Lifetime $\tX$ & $3.26\,$s\\
 $(B_{ee},B_{\nu_e})$ & $(0.52,0.48)$\\
 $\neff$, $\yp$, $\Dfive$ & $2.98$, $0.2447$, $2.450$\\
 $\Delta\chi^2$, Eq.~\eqref{eq:joint-region} & $1.9$\\\hline
 \end{tabular}
 \caption{Parameters and predictions of the CP-odd realization of mixed decays. The pseudoscalar $S$ mixes with $X$ and gives it the electron coupling during reheating; the vectorlike lepton and the electroweak triplet generate the couplings of both scalars.}
 \label{tab:sm-cpodd}
\end{table}

In one explicit realization (Table~\ref{tab:sm-cpodd}), a CP-odd $X$ couples to electron neutrinos directly and to electrons through its mass mixing with a heavier pseudoscalar $S$, which has an electron Yukawa coupling. The mixing depends on a modulus that rolls after reheating, so the electron coupling of $X$ then falls below the stellar limits, while its neutrino coupling stays constant. A first-generation vectorlike lepton, an $SU(2)_L$ singlet, generates these interactions together with an electroweak triplet $\Delta$. Their masses exceed the ATLAS lower limits of $320\,\mathrm{GeV}$ on such fermions~\cite{ATLAS:2024mrr} and about $1.03\,\mathrm{TeV}$ on the doubly charged member of $\Delta$~\cite{ATLAS:2022pbd}. A discrete lepton symmetry forbids the coupling $HH\Delta^\dagger$, through which triplet exchange would generate the Weinberg operator $(L\cdot H)(L\cdot H)$, so $\Delta$ acquires no vacuum expectation value. The decays of $X$ reheat the plasma to $\Trh\simeq0.39\,\MeV$, with a pattern within the flavor families of Fig.~\ref{fig:minima}. The full calculation, with the exponent $3/2$ in Eq.~\eqref{eq:sm-ee-medium} replaced by $1/2$ for a CP-odd coupling, places this history inside the joint region. The pseudoscalar $S$ also evades laboratory bounds: it decays into $e^+e^-$ with $c\tau_S\simeq0.6\,$mm, so those produced in the E137 experiment do not reach its detector~\cite{Bjorken:1988as}. The Orsay and KEK beam dumps and NA64 exclude only lighter states~\cite{Davier:1989wz,Konaka:1986cb,NA64:2021aiq}, and the SINDRUM search for $\pi^+\to e^+\nu_ee^+e^-$ constrains only couplings at least five times larger~\cite{SINDRUM:1989qan,Altmannshofer:2022ckw}. This mediator lies within the projected reach of LDMX~\cite{Berlin:2018bsc,Akesson:2022vza,Gaiser:2026kkx}, which would test this realization.

The reheater can also consist of non-topological solitons such as Q-balls~\cite{Coleman:1985ki}, which can dominate the Universe before BBN and release their energy into leptons~\cite{Kawasaki:2002hq,Kawasaki:2022hvx}. The released particles carry energies set by an internal oscillation frequency, which can lie far below the mass of the free quanta of the field. Stars cannot assemble such objects and emit only single constituents, whose production falls steeply when they are much heavier than the stellar temperature. A viable reheater of this kind must keep its frequency constant while it shrinks, stay coherent despite its interactions with leptons, and reach the required abundance before it releases its energy.

For example, solitons of two complex fields, with an internal rotation at several MeV and free quanta at the GeV scale, can emit $e^+e^-$ and $\nu_e\bar\nu_e$ pairs through couplings of order $10^{-11}$. Their energy release might then follow that of the decays in Fig.~\ref{fig:minima} near $\Trh\simeq0.4\,\MeV$ and satisfy the joint BBN and CMB constraint of Eq.~\eqref{eq:joint-region}. These couplings can arise from a heavy electroweak triplet, as in the CP-odd example. It then also contributes to the scattering of electrons off positrons and off one another, with an effect that can lie just below the present limits from LEP and E158~\cite{ALEPH:2013dgf,SLACE158:2005uay}. The MOLLER experiment could detect it~\cite{MOLLER:2014iki}.

\section{Secondary pions in electromagnetic reheating}
\label{sec:sm-pions}
Subdominant decays producing charged pions can strengthen reheating bounds because pion absorption turns protons into neutrons. We calculate the pion production accompanying decays into electrons or photons, then include these proton-to-neutron conversions in the nuclear reaction network while holding the electromagnetic plasma evolution fixed.

\begin{figure}[t!]
 \centering
 \includegraphics[width=0.45\textwidth]{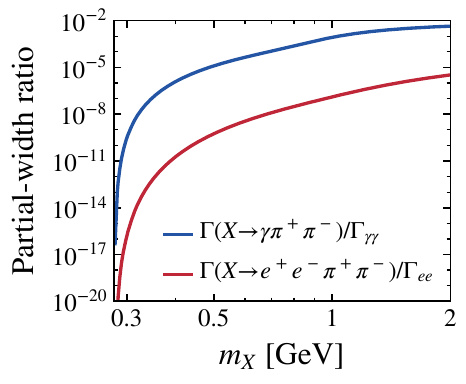}\hfill
 \includegraphics[width=0.52\textwidth]{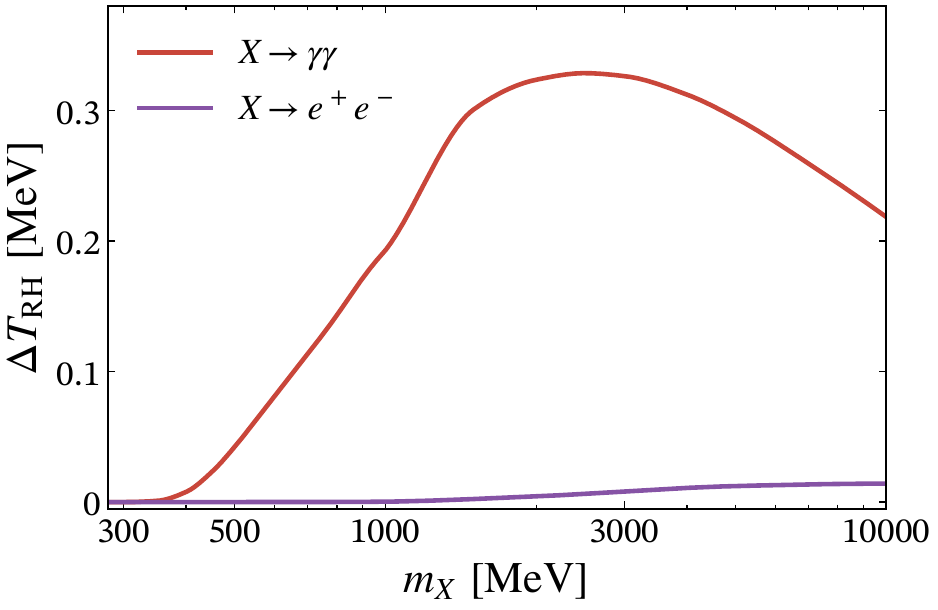}
 \caption{Radiative pions and their effect on reheating. Left: dipion widths relative to the leading photon (blue) or electron (red) decay, Eqs.~\eqref{eq:eepipi} and~\eqref{eq:photonpions}. Right: the increase of the minimum temperature, Eq.~\eqref{eq:pion-response}, for the inclusive photon-induced yield (red) and radiation from electrons (purple), at fixed plasma evolution.}
 \label{fig:pions}\label{fig:pion-response}
\end{figure}

The dipion spectral function $R_{\pi\pi}(s)$ relates pion production through a virtual photon to the measured annihilation cross section $\sigma(e^+e^-\to\pi^+\pi^-)$. For dipion invariant mass squared $s$, fine-structure constant $\alpha$, charged pion mass $m_{\pi^\pm}$, and pion form factor $F_\pi(s)$, the corresponding differential widths are
\begin{align}
 R_{\pi\pi}(s)&=\frac{\sigma(e^+e^-\to\pi^+\pi^-)}{4\pi\alpha^2/(3s)}
 =\frac14\left(1-\frac{4m_{\pi^\pm}^2}{s}\right)^{3/2}|F_\pi(s)|^2,
 \label{eq:rpipi}\\
 \frac{d\Gamma(X\to e^+e^-\pi^+\pi^-)}{ds}
 &=\Gamma[X\to e^+e^-\gamma^*(\sqrt{s})]\frac{\alpha R_{\pi\pi}(s)}{3\pi s},
 \label{eq:eepipi}\\
 \frac{d}{ds}\frac{\Gamma(X\to\gamma\pi^+\pi^-)}{\Gamma(X\to\gamma\gamma)}
 &=\frac{2\alpha R_{\pi\pi}(s)}{3\pi s}\left(1-\frac{s}{m_X^2}\right)^3.
 \label{eq:photonpions}
\end{align}
Here, $\Gamma[X\to e^+e^-\gamma^*(\sqrt{s})]$ is the width for emitting a virtual photon of mass $\sqrt{s}$ in an electron-pair decay. We use the pion form factor of Ref.~\cite{Ilten:2018crw}. For Eq.~\eqref{eq:eepipi}, we include radiation from both charged leptons and their interference, all polarizations of $\gamma^*$, and the exact electron mass, with $4m_{\pi^\pm}^2\leq s\leq(m_X-2m_e)^2$. Equation~\eqref{eq:photonpions} applies to local CP-even or CP-odd couplings of $X$ to photons~\cite{EscuderoAbenza:2025tsi}. A momentum-dependent transition form factor $F_X(s)$ for $X\gamma\gamma^*$ multiplies that expression by $|F_X(s)/F_X(0)|^2$. Equations~\eqref{eq:eepipi} and~\eqref{eq:photonpions} give probabilities of dipion production of order $\alpha^2$ and $\alpha$ relative to the leading electron and photon decay widths, respectively (Fig.~\ref{fig:pions}).

The branching fraction for producing a charged pion pair is $\mathcal B_{\pi\pi}=B_{ee}r_{e\pi\pi}+B_{\gamma\gamma}r_{\gamma\pi\pi}$. Here, $B_{\gamma\gamma}$ is the vacuum two-photon branching fraction, $r_{e\pi\pi}=\Gamma(X\to e^+e^-\pi^+\pi^-)/\Gamma(X\to e^+e^-)$, and $r_{\gamma\pi\pi}=\Gamma(X\to\gamma\pi^+\pi^-)/\Gamma(X\to\gamma\gamma)$. Each of these decays produces one pion of each charge. For $B_{\gamma\gamma}=1$, we also include prompt pions from other hadronic final states, using the mean number of negative pions per $X$ decay, $\overline N_{\pi^-}$, calculated in Ref.~\cite{EscuderoAbenza:2025tsi}. The production rates per physical volume and time, $S_{\pi^\pm}$, obey
\begin{equation}
 S_{\pi^+}+S_{\pi^-}=\frac{2\overline N_{\pi^-}}{m_X}\frac{\rho_X}{\tau_X}.
 \label{eq:mean-pion-source}
\end{equation}
We assume that charged pions lose their kinetic energy before reacting. We then evolve their absorption, decay, and annihilation together with neutron-proton conversion and nuclear reactions~\cite{Reno:1987qw,Kohri:2001jx,Hasegawa:2019jsa}. The large number of reheaters per baryon makes even a small pion yield important. At sufficiently early times, however, rapid weak reactions restore the neutron-to-proton ratio after pion-induced conversions. Accordingly, varying the initial neutron fraction over its full physical range changes its value at the onset of nuclear burning by less than $2\times10^{-8}$.

For leading electromagnetic channel $c=ee$ or $\gamma\gamma$, the shift of the lower reheating bound due to pion production is
\begin{equation}
 \Delta\Trh^{(c)}(m_X)=\Trhlow^{(c)}(m_X;S_{\pi^\pm})-\Trhlow^{(c)}(m_X;0),
 \label{eq:pion-response}
\end{equation}
where the second term omits pion production. We vary the baryon density to maximize the likelihood, use standard BBN as the common best-fit reference, and hold the electromagnetic plasma and neutrino evolution fixed. Electrons and photons from heavy reheater decays thermalize rapidly. Below $T=5\,\MeV$, their occupation numbers at the injection momentum $m_X/2$ are negligible, so we match the energy density remaining in $X$ at that temperature between the electron and photon cases.

Pion production raises the reheating bound more strongly for photon decays. The correction for electron decays is too small to see on the scale of Fig.~\ref{fig:minima}. For photon decays, the increase peaks at a few GeV. The shift follows the pion source per unit of injected energy, $(S_{\pi^+}+S_{\pi^-})\tau_X/\rho_X=2\overline N_{\pi^-}/m_X$, from Eq.~\eqref{eq:mean-pion-source}. The factor $(1-s/m_X^2)^3$ in Eq.~\eqref{eq:photonpions} reduces the production of hadrons whose invariant mass $\sqrt s$ approaches $m_X$. Up to a few GeV, raising $m_X$ removes this reduction for the $\rho$ resonance and multipion states, so $\overline N_{\pi^-}$ rises faster than $m_X$. For heavier reheaters, $\overline N_{\pi^-}$ grows more slowly than $m_X$, and the source per unit energy falls. However, the pion calculation extends only to hadronic invariant masses of $2\,\mathrm{GeV}$. Adding heavier final states, kaon and nucleon reactions, direct quark or gluon couplings, or the effect of these reactions on plasma evolution can strengthen the bound. The peak therefore depends on the hadronic channels included.

\section{Neutrino masses and relic abundances}
\label{sec:sm-relics}
Late reheating changes the mass density of active neutrinos, suppresses sterile production, and dilutes earlier relics. The following calculations connect these consequences to the compatible decay patterns in Table~\ref{tab:examples}.

\subsection{Active neutrinos and matter clustering}
Nearly standard radiation density can coexist with far fewer massive neutrinos when each particle carries more energy~\cite{Oldengott:2019mass,Alvey:2021distribution}. After collisions cease, we calculate the number and energy densities of the neutrino mass eigenstates using the phase-averaged oscillation prescription. The standard reference for one species, including antineutrinos, is $n_{\nu,\rm std}=3\zeta(3)T_{\nu,\rm std}^3/(2\pi^2)$ and $\rho_{\nu,\rm std}=7\pi^2T_{\nu,\rm std}^4/120$, where $T_{\nu,\rm std}=(4/11)^{1/3}T$ and $\zeta$ is the Riemann zeta function. For each state, $f_i(q)$ is the occupation per helicity, with $q=p/T_{\nu,\rm std}$ and equal neutrino and antineutrino distributions. We define the combined number density $n_i$, relativistic energy density $\rho_i^{\rm rel}$, and their moments by
\begin{equation}
 \begin{gathered}
 n_i=\frac{T_{\nu,\rm std}^3}{\pi^2}\int d\ln q\,q^3 f_i(q),\qquad
 \rho_i^{\rm rel}=\frac{T_{\nu,\rm std}^4}{\pi^2}\int d\ln q\,q^4 f_i(q),\\
 r_i=\frac{n_i}{n_{\nu,\rm std}},\qquad
 \epsilon_i=\frac{\rho_i^{\rm rel}}{\rho_{\nu,\rm std}},\qquad
 \bar q_i=\frac{\langle p\rangle_i}{T_{\nu,\rm std}}=\frac{\rho_i^{\rm rel}}{n_iT_{\nu,\rm std}}.
 \end{gathered}
 \label{eq:sm-active-moments}
\end{equation}
These measure the relative number, energy, and mean momentum. With no dark radiation, $\sum_i\epsilon_i=\neff$. The present non-relativistic neutrino mass density, expressed as a fraction of the critical density, obeys $\Omega_\nu^{\rm NR}h^2\simeq M_\rho/(94.1\,\mathrm{eV})$, where $M_\rho\equiv\sum_i m_ir_i$ equals the mass sum for thermal neutrinos.

\begin{figure}[t!]
 \centering
 \includegraphics[width=0.45\textwidth]{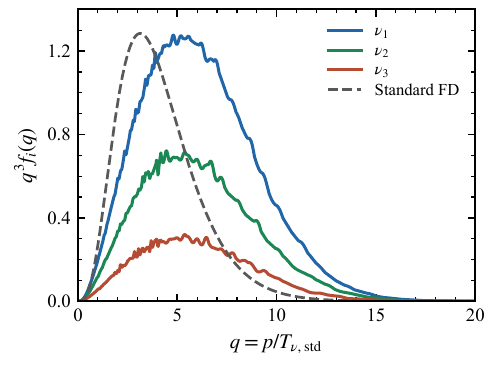}
 \caption{Cosmic neutrino background for the compatible $m_X=6\,\MeV$, $\Trh=0.42\,\MeV$ decay pattern of Fig.~\ref{fig:active-mass}. Final occupations of mass eigenstates $f_i(q)$ per helicity are compared with the standard Fermi-Dirac form $(e^q+1)^{-1}$, where $q=p/T_{\nu,\rm std}$ and $p$ is the neutrino momentum in the cosmological rest frame. The ordinate is proportional to particle number per logarithmic momentum interval, with the absolute normalization of Eq.~\eqref{eq:sm-active-moments}. Curves show means of three realizations.}
 \label{fig:active-moments}
\end{figure}

The $m_X=6\,\MeV$, $\Trh=0.42\,\MeV$ example depletes mainly the heaviest state (Fig.~\ref{fig:active-moments} and Table~\ref{tab:active-moments}). For normal ordering and $m_1=0$, $M_\rho$ therefore falls to about one third of the physical mass sum, a larger reduction than that of the total neutrino number. The remaining neutrinos carry more energy than thermal ones, with mean momenta well above the Fermi-Dirac value $3.15T_{\nu,\rm std}$. Another compatible history reduces $M_\rho$ further: for $(B_{\nu_e},B_{\nu_\mu},B_{\nu_\tau})/B_\nu=(0.92,0.06,0.02)$ and $B_{ee}\simeq0.50$, at $m_X=11\,\MeV$ and $\Trh=0.58\,\MeV$ with $\Delta\chi^2\simeq4.6$, it is $0.011\,$eV, about a fifth of the mass sum.

For inverted ordering, each eigenstate keeps its flavor content, and only the thermal matter potential changes the evolution. It is negligible below about $3\,\MeV$, while at higher temperatures the neutrinos present make up less than $10^{-4}$ of their final number. We therefore use the same $r_i$, now with $m_3=0$. The splittings of Eq.~\eqref{eq:sm-mixing} give $m_1\simeq0.0506\,$eV and $m_2\simeq0.0513\,$eV. Since the most depleted state $\nu_3$ is now massless, $M_\rho\simeq0.088\,$eV, about $87\%$ of the mass sum and above the values preferred by fits combining CMB and DESI measurements~\cite{DESI:2025DR2}.

\begin{table}[t!]
 \centering
 \begin{tabular}{ccccc}
 \hline
 $i$ & $m_i/\mathrm{eV}$ & $r_i$ & $\epsilon_i$ & $\bar q_i$\\\hline
 1 & 0 & 1.110 & 1.610 & 4.570\\
 2 & 0.00858 & 0.626 & 0.894 & 4.500\\
 3 & 0.05060 & 0.275 & 0.387 & 4.433\\\hline
 \end{tabular}
 \caption{Masses and relic neutrino moments for the compatible $m_X=6\,\MeV$, $\Trh=0.42\,\MeV$ decay pattern of Fig.~\ref{fig:active-mass}, with normal ordering and $m_1=0$. Relative number densities $r_i$, relativistic energy densities $\epsilon_i$, and mean momenta $\bar q_i$ follow Eq.~\eqref{eq:sm-active-moments} and are averaged over ten realizations. The mass sum is $0.0592\,\mathrm{eV}$ and $M_\rho=0.0193\,\mathrm{eV}$; its standard error is $3\times10^{-5}\,\mathrm{eV}$.}
 \label{tab:active-moments}
\end{table}

We calculate linear matter and CMB lensing power with \textsc{Class}~\cite{Blas:2011rf,Lesgourgues:2011rh}, using the final mass eigenstate distributions and adiabatic primordial perturbations. The Fermi-Dirac comparison has the same physical masses and relativistic energy density for each state, which fixes its temperature to $T_i/T_{\nu,\rm std}=\epsilon_i^{1/4}$. Both models have $\omega_b=0.0224$, total physical matter density $\omega_m=\Omega_mh^2=0.143$, $h=0.674$, and $Y_{\rm CMB}=\mathcal Y(\yp)\simeq0.2449$. Here, $h$ is the Hubble constant in units of $100\,\mathrm{km\,s^{-1}\,Mpc^{-1}}$ and $\Omega_m$ is today's total matter density as a fraction of the critical density. We adjust the cold dark matter density to compensate for the different neutrino mass densities. The massless neutrino state contributes the radiation density obtained from its distribution.

The reduced mass density raises linear matter power by $\simeq1\%$ at $k=0.1$ to $1\,h\,\mathrm{Mpc}^{-1}$ and CMB lensing power by up to $0.7\%$ for angular multipoles $\ell\leq2000$. Multiplying Fermi-Dirac distributions by a constant dilution factor and adjusting their temperatures to match both number and energy densities leaves matter power differences below $0.014\%$ over $k=0.001$ to $5\,h\,\mathrm{Mpc}^{-1}$ at redshifts zero and three. Refining spectral reconstruction and integration changes matter power and lensing by less than $0.0002\%$. Neutrino number and energy densities therefore account for nearly all of this example's effect. We leave cosmological mass bounds from fits with these distributions to future work.

The CMB likelihood of Sec.~\ref{sec:sm-bbn} treats the relic neutrinos as thermal, with a single massive state of $0.06\,$eV. We test this assumption for histories close to the boundary of the joint region: near the lowest allowed reheating temperatures at $m_X=5$, $6$, and $9\,\MeV$, including the $\Trh=0.42\,\MeV$ example and a pattern with $B_{\nu_e}=B_\nu$, the two rows of Table~\ref{tab:examples} at $\Trh=1.05\,\MeV$, and decays mainly into $\nu_\mu$ or $\nu_\tau$, which enrich $\nu_3$. For each of them, Fermi-Dirac shapes reproduce the calculated number and energy densities of the mass eigenstates, for normal ordering with $m_1=0$. At fixed $\omega_b$, $\neff$, and $Y_{\rm CMB}$, we refit five cosmological and four calibration parameters to the Planck, ACT, and SPT data~\cite{SPT-3G:2025bzu}, using a linear expansion of the observables around each history.

The CMB $\chi^2$ then changes by $-0.30$ to $+0.51$ and grows with $M_\rho$, which spans $0.006$ to $0.11\,$eV in these histories. Removing the neutrino mass density entirely lowers $\chi^2$ by $0.33$, and replacing the single $0.06\,$eV state by thermal eigenstates with normal ordering shifts it by only $0.001$. Refining the numerical CMB spectra changes these values by at most $0.06$. Near the lowest reheating temperatures, the decays favor electron neutrinos, and $\chi^2$ falls by $0.1$ to $0.3$. The joint $\chi^2$ there grows by several hundred per unit of $\ln\Trh$, mostly through helium, so the bound moves by less than $0.001\,\MeV$. The regions and limits of Fig.~\ref{fig:minima} therefore use the thermal likelihood of Sec.~\ref{sec:sm-bbn}.

\subsection{Sterile neutrinos within experimental reach}
Near-MeV reheating suppresses sterile neutrino production at mixings accessible to laboratory searches~\cite{Gelmini:2004ah,Yaguna:2007wi,Hasegawa:2020ctq}. The mixing matrix relates flavor and mass eigenstates through $\nu_\alpha=\sum_{i=1}^4U_{\alpha i}\nu_i$. In the short-baseline limit, the disappearance amplitude is $\sin^2(2\theta_{\alpha4})=4|U_{\alpha4}|^2(1-|U_{\alpha4}|^2)$, where $\alpha=e,\mu,\tau$.

We take a stable state with mass $m_4=23.4\,\mathrm{eV}$, $|U_{\mu4}|^2=10^{-3}$ or $4.4\times10^{-3}$, $U_{e4}=U_{\tau4}=0$, no extra interactions or primordial lepton asymmetry, and negligible initial occupation at $20\,\MeV$. These mixings give $\sin^2(2\theta_{\mu4})\simeq0.0040$ and $0.0175$, respectively. For reheater number density $n_X$, the dilute source of $\nu_4$ is $2\Gamma_Xn_X\sum_\alpha B_{\nu_\alpha}|U_{\alpha4}|^2$ per volume and time. The $12.5\,\MeV$, $\Trh=1.05\,\MeV$ decay pattern in Table~\ref{tab:examples} has $B_{\nu_\mu}=0$, so $X$ creates no $\nu_4$ directly.

We calculate sterile neutrino production by weak reactions to leading order in $|U_{\mu4}|^2$, using the active neutrino spectra and expansion rate from the three-neutrino calculation. The production rates include interference between scattering amplitudes and Pauli blocking by active neutrinos. Most sterile neutrinos are produced below $T=5\,\MeV$, where the plasma's refractive potential has a negligible effect on mixing. Production above $5\,\MeV$ adds less than one percent, and Pauli blocking by the produced states changes the rate by less than $0.04\%$. The sterile energy density stays below $0.3\%$ of the total; its effect on active neutrino evolution and nuclear abundances is neglected. Once produced, the sterile states also interact through their $\nu_\mu$ admixture, at a level suppressed by $|U_{\mu4}|^2$, and each such collision can convert them into $\nu_\mu$. An upper bound on the scattering rate, integrated over the reheating history, limits the affected fraction to $6\%$ for $|U_{\mu4}|^2=4.4\times10^{-3}$ and $1.3\%$ for $10^{-3}$, so we neglect this loss. After collisions cease, vacuum oscillations preserve the mass eigenstate occupations, so the suppressed sterile population persists.

For sterile number density $n_4$ including both helicities and relativistic energy density $\rho_4^{\rm rel}$, the relative abundance $r_N$, extra radiation $\Delta N_4$, and present mass density $\Omega_4$ in units of the critical density are
\begin{equation}
 r_N=\frac{n_4}{n_{\nu,\rm std}},\qquad
 \Delta N_4=\frac{\rho_4^{\rm rel}}{\rho_{\nu,\rm std}},\qquad
 \Omega_4h^2=\frac{m_4}{94.1\,\mathrm{eV}}r_N.
 \label{eq:sm-sterile-moments}
\end{equation}
The standard densities refer to one active species including antineutrinos, as defined above.
\begin{table}[t!]
 \centering
 \begin{tabular}{cccc}
 $|U_{\mu4}|^2$ & $r_N$ & $\Delta N_4$ & $\Omega_4h^2$\\\hline
 $10^{-3}$ & $(7.87\pm0.01)\times10^{-4}$ & $(7.96\pm0.01)\times10^{-4}$ & $(1.957\pm0.003)\times10^{-4}$\\
 $4.4\times10^{-3}$ & $(3.461\pm0.005)\times10^{-3}$ & $(3.504\pm0.006)\times10^{-3}$ & $(8.61\pm0.01)\times10^{-4}$
 \end{tabular}
 \caption{Sterile abundances defined by Eq.~\eqref{eq:sm-sterile-moments} for the $m_X=12.5\,\MeV$, $\Trh=1.05\,\MeV$ decay pattern. Means and Monte Carlo standard errors use ten independent realizations of the active neutrino distributions. $\Delta N_4$ measures the sterile contribution at fixed plasma and active spectra.}
 \label{tab:sterile-reheating}
\end{table}

\begin{figure}[t!]
 \centering
 \includegraphics[width=0.40\textwidth]{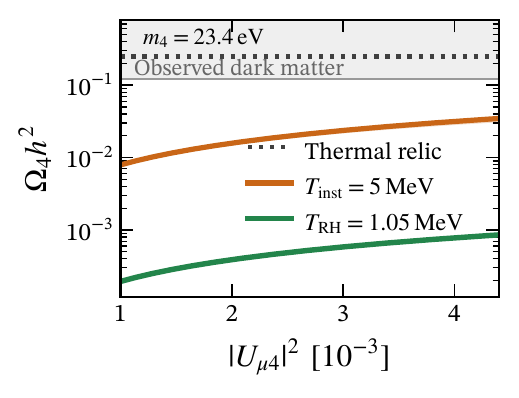}\hspace{0.03\textwidth}
 \includegraphics[width=0.40\textwidth]{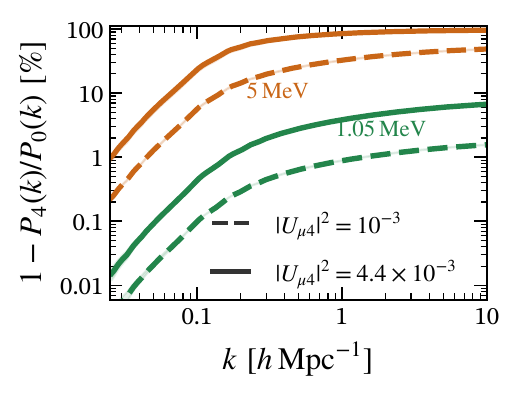}
 \caption{Sterile relic density and linear matter power suppression for $m_4=23.4\,\mathrm{eV}$ and muon mixing. Left: mass density, Eq.~\eqref{eq:sm-sterile-moments}, for the $m_X=12.5\,\MeV$, $\Trh=1.05\,\MeV$ decay pattern in Table~\ref{tab:examples} (green), instantaneous reheating to $5\,\MeV$ (orange), and a thermal relic (dotted); the gray region lies above the dark matter density~\cite{Planck:2018nkj}. Right: power suppression today, Eq.~\eqref{eq:sterile-power}. The orange band spans the dilute and saturating estimates of Eq.~\eqref{eq:sterile-five-mev}; line styles distinguish the indicated mixings.}
 \label{fig:sterile-relics}
\end{figure}

To quantify the suppression, we compare with instantaneous reheating to $T_{\rm inst}=5\,\MeV$ and an equilibrium active neutrino bath. Writing $q=p/T_{\nu,\rm std}$ and $f_{\rm FD}(q)=(e^q+1)^{-1}$, the sterile occupation in the dilute limit and in an approximation that saturates production as equilibrium is approached are~\cite{Gelmini:2004ah}
\begin{equation}
 f_4^{\rm dil}(q)=Aqf_{\rm FD}(q),\qquad
 f_4^{\rm sat}(q)=(1-e^{-Aq})f_{\rm FD}(q),\qquad
 A=3.2d_\mu\sin^2(2\theta_{\mu4})\left(\frac{T_{\rm inst}}{5\,\MeV}\right)^3,
 \label{eq:sterile-five-mev}
\end{equation}
where $d_\mu=0.79$ is the dimensionless coefficient for weak production through mixing with muon neutrinos. This comparison assumes negligible suppression of mixing by the plasma and starts sterile production at $5\,\MeV$. Reheating through decays into electrons and neutrinos reduces the sterile abundance by a factor of $\simeq40$ (Fig.~\ref{fig:sterile-relics}).

Using \textsc{Class}, we compare the linear matter power spectrum $P_4$ with the sterile population to $P_0$ with its mass density replaced by cold dark matter. The fractional suppression today is
\begin{equation}
 S(k)=1-\frac{P_4(k,z=0)}{P_0(k,z=0)}.
 \label{eq:sterile-power}
\end{equation}
We hold the primordial perturbations, active neutrino radiation density, and total matter density fixed, with $h=0.674$, $\omega_b=0.02237$, cold dark matter density $\omega_{\rm cdm}=\Omega_{\rm cdm}h^2=0.12$ in $P_0$, and an active neutrino state of mass $0.06\,\mathrm{eV}$. The active neutrino spectra are approximated by thermal distributions with the calculated $\neff$ for mixed reheating and $\neff=3.044$ for instantaneous reheating. Thus, the change in power measures the sterile neutrino contribution; the effect of non-thermal active spectra is calculated in the preceding subsection.

At the larger mixing, the instantaneous $5\,\MeV$ scenario produces too much hot dark matter even when production saturates toward equilibrium (Fig.~\ref{fig:sterile-relics}, left). A thermal relic with the same relativistic energy density and late mass density has a mass near $12\,\mathrm{eV}$ and accounts for about $27\%$ of the dark matter, exceeding the Planck 95\% limit of $8\%$~\cite{Peters:2023asu}. Its matter power spectrum differs from that of the instantaneous scenario by less than $1.5\%$. Reheating through electron and neutrino decays lowers the sterile fraction to about $0.7\%$ and leaves percent-level suppression of matter power (Fig.~\ref{fig:sterile-relics}, right). The smaller mixing gives a lower abundance and weaker suppression. We leave detailed confidence regions, including coupled active and sterile neutrino evolution, to future work.

The larger mixing is also within the projected sensitivity of muon neutrino disappearance searches, at $\Delta m_{41}^2\simeq550\,\mathrm{eV}^2$ and $\sin^2(2\theta_{\mu4})\simeq0.018$. For five years of data, nuSCOPE reaches $|U_{\mu4}|^2\simeq0.002$ at two standard deviations using the spectral shape and neglecting detector systematic uncertainties~\cite{Blanchet:2026jwx}. SHiP would probe approximately $0.0043$ or $0.0018$ at 90\% confidence if its proposed far detector records $0.1$ or $1$ times as many events as the near detector, respectively~\cite{Choi:2025vby}. These sensitivity estimates assume correlated normalization errors. Additional detector uncertainties can weaken the reach.

\subsection{Dilution of an axion-like condensate}
Reheating dilutes the energy density of a decoupled, oscillating condensate relative to the photon entropy~\cite{Visinelli:2009kt,Blinov:2019jqc}. A homogeneous axion-like field $\phi_a(t)$ of constant mass $m_a$ obeys $\ddot\phi_a+3H\dot\phi_a+m_a^2\phi_a=0$, with dots denoting time derivatives. Initially, $\phi_a(t_i)=\phi_{a,i}$, $\dot\phi_a(t_i)=0$, and $m_a\ll H(t_i)$, so Hubble friction holds the field nearly fixed. Once oscillations begin, their averaged energy density redshifts as $a^{-3}$. We calculate its ratio to photon entropy density after electron-positron annihilation to obtain the present condensate abundance.

For $m_a=3\times10^{-14}\,\mathrm{eV}$, the $m_X=11.25\,\MeV$, $\Trh=1.05\,\MeV$ pattern in Table~\ref{tab:examples} dilutes the condensate four times more than the electron-only history in its last row, at fixed field parameters. In the mixed case, $\phi_a$ starts to oscillate while $X$ still dominates the expansion. Both energy densities then fall as $a^{-3}$ until the decays raise the comoving entropy about thirtyfold. The final condensate density relative to photons is therefore set by $\Trh$ and $\phi_{a,i}$ and does not depend on the initial amount of $X$, provided that $X$ dominates when the oscillations begin. In a fluid model of this history, dividing the reheater abundance at $20\,\MeV$ by $10^4$ or multiplying it by ten changes $\Omega_ah^2$ by less than $0.01\%$. When electrons receive all decays, at $\Trh=5.4\,\MeV$, most of the energy is released before the onset, and the comoving entropy grows afterwards by only $20\%$. In the mixed case, the oscillations begin in a colder plasma, which offsets part of the difference. For the mixed decays, an initial amplitude $\phi_{a,i}\simeq2.4\times10^{15}\,\mathrm{GeV}$ gives the observed dark matter density, $\Omega_a h^2\simeq0.12$. The abundance scales as $\phi_{a,i}^2$, so a smaller amplitude suffices under electromagnetic reheating. The field contributes less than $4\times10^{-5}$ of the total energy density during BBN. Constraints on primordial isocurvature perturbations depend on how the condensate was produced.

\end{document}